\documentclass[12pt]{article}
\usepackage[english]{babel}
\usepackage[utf8]{inputenc}
\usepackage[T1]{fontenc}
\usepackage{authblk}
\usepackage{graphicx}
\usepackage{amsmath,mathrsfs,dsfont}
\usepackage{amsthm}
\usepackage[all]{xy}
\usepackage{graphics}
\usepackage{tensor}
\usepackage{verbatim}
\usepackage{nicefrac}
\usepackage{amsfonts}
\usepackage{amssymb,latexsym}
\usepackage{color}
\usepackage{mathrsfs,dsfont}
\usepackage{appendix}
\usepackage{slashed}
\usepackage{cite}
\usepackage{comment}
\definecolor{verde}{cmyk}{.83,.21,1,.08}

\DeclareMathOperator{\ii}{\mathrm{i}}
\DeclareMathOperator{\ee}{\mathrm{e}}

\begin{document}

\title{
Dimensional Deception from Noncommutative Tori:\\{\Large An alternative to Ho\v{r}ava-Lifschitz}}
\author[1,2,3]{Fedele Lizzi\thanks{fedele.lizzi@na.infn.it}}
\author[4,5]{Aleksandr Pinzul\thanks{apinzul@unb.br}}
\affil[1]{INFN, Sezione di Napoli, Complesso Univ. Monte S. Angelo, Via Cintia, I-80126 Napoli, Italy}
\affil[2]{Dipartimento di Fisica ``E.\ Pancini'', Universit\`{a} di Napoli “Federico II”, Complesso Univ. Monte S. Angelo, Via Cintia, I-80126 Napoli, Italy}
\affil[3]{Departament de F\'{\i}sica Qu\`antica i Astrof\'{\i}sica and Institut de C\'{\i}encies del Cosmos (ICCUB),
Universitat de Barcelona, Mart\'\i i Franqu\'es 1, 08028 Barcelona, Spain}
\affil[4]{Universidade de Bras\'{\i}lia Instituto de F\'{\i}sica 70910-900, Bras\'{\i}lia, DF, Brasil}
\affil[5]{International Center of Physics C.P. 04667, Brasilia, DF, Brazil}
\date{}
\maketitle

\begin{abstract}
We study the dimensional aspect of the geometry of quantum spaces. Introducing a physically motivated notion of the scaling dimension, we study in detail the model based on a fuzzy torus. We show that for a natural choice of a deformed Laplace operator, this model demonstrates quite non-trivial behaviour: the scaling dimension flows from 2 in IR to 1 in UV. Unlike another model with the similar property, the so-called Horava-Lifshitz model, our construction does not have any preferred direction. The dimension flow is rather achieved by a rearrangement of the degrees of freedom. In this respect the number of dimensions is deceptive. Some physical consequences are discussed.
\end{abstract}

\section{Introduction}

The usual Riemannian geometry on a compact space is completely specified by some algebraic data, the so-called spectral triple. It was shown \cite{Connes:2008vs} that if this data satisfies some natural conditions then there is one-to-one correspondence between the algebraic and geometric formulations of a compact geometry. The main ingredients of this data are: 1) a $C^*$-algebra, $\mathcal{A}$; 2) Dirac operator, $\mathcal{D}$; 3) Hilbert space, $\mathcal{H}$, on which $\mathcal{A}$ and $\mathcal{D}$ are represented. The advantage of the algebraic formulation is that it admits non-trivial generalizations which lead to different non-classical geometries \cite{Connes:1994yd}. In this paper we are interested in just one aspect of such geometries - their spectral dimension. There is a classical result due to Weyl, which we review below, that states that the usual dimension of a compact geometry can be inferred from the growth rate of the eigenvalues of the corresponding Laplacian. On the example of several geometries related to a 2-dimensional torus, we will illustrate the passage from the classical commutative geometry to highly noncommutative one, as well as demonstrate the non-trivial behaviour of the spectral dimension given by the physically relevant generalization of the Weyl's result. For this we will not need the whole algebraic (spectral) data but only the information on the spectrum of the Dirac operator (actually, related to it Laplace operator or its generalization). The (generalized) geometries that we will use in this work are
\begin{enumerate}
  \item A commutative 2d torus, $\mathbb{T}^2$;
  \item A noncommutative 2d torus, $\mathbb{T}^2_\theta$;
  \item A torus with one commutative and one fuzzy dimension;
  \item A fuzzy torus, $\mathbb{T}^2_n$ (this is the main object of our work).
\end{enumerate}
Here and in the following by \emph{fuzzy} spaces (for a review see~\cite{Balachandran:2005ew}) we mean finite matrix approximations to a space, commutative or otherwise. Fuzzy spaces are known to have interesting and novel featurs in field theory, like the appearance of different phases (striped phases) in the ultraviolet~\cite{Gubser:2000cd, Panero:2006bx, GarciaFlores:2009hf,Lizzi:2012xy}. We will work in two dimensions for definitiveness and to avoid the proliferation of indices, but the main results can be stated in higher dimensions, and we will comment on this in the conclusions. We will see that with a suitable generalization of Weyl's spectral dimension we will have that in our model the number of dimensions scales with energy from two to one, but in principle could be much more complicated, which we will demonstrate on a simple example of a torus with one commutative and one fuzzy dimension. At the same time the fundamental isometries of the model are always unbroken, all along there is no preferred direction. The ``high energy'' lower dimensional space is highly noncommutative.  In particular our model appears as a 2d torus at large scales/low energy, and as the direct sum of two one-dimensional circles at short scales/high energy.  The reasons for which the number of dimensions change at different scales is very different with respect to more conventional mechanisms, such as, e.g., the Horava-Lifschitz one\cite{Horava:2009uw,Horava:2009if}. Usually the number of dimensions changes by selecting one particular dimension (say time) and modifying the Laplacian to have this dimension behave differently from the others, so to have a spectral flow alter the dimensions. In our case the dimensions are treated always on an equal footing. While the original space is highly noncommutative, i.e. it is genuinely ``quantum'' at short distances, at low energy (long distances) the deceptively higher number of dimensions \emph{emerges} from a rearrangement of the eigenvalues and eigenvectors of the Laplacian (or the Dirac operator) which simulate a higher dimensional commutative space. Such a picture is an explicit realization of the situation when the microscopic (UV, ``fundamental'') degrees of freedom are completely different from the macroscopic (IR, ``effective'') ones. Although here we present a particular model, the fundamental ideas are more general and can in principle be applied to more realistic spaces, even if the technical difficulties can grow.

The organization of our paper as follows. In section \ref{spectvsscal} we discuss some natural generalization of the notion of a spectral dimension. Then, using this generalization, we illustrate our main idea on several, gradually more and more noncommutative, examples. Section \ref{se:approximation} is a brief review of a specific matrix approximation of a noncommutative torus due to Elliot and Evans. In section \ref{se:truncation} this approximation is realized as a particular truncation of the algebra of a noncommutative torus. After this, in section \ref{se:derivatives}, we discuss one of the natural choices for the deformed derivations of this matrix algebra as well as calculate the spectrum of the associated (deformed) Laplacian. Finally, in section \ref{se:spectral}, we analyze the spectral dimension of our model in two limiting cases, UV and IR. We conclude with some discussion of our results and possible future developments. Because the Elliot-Evans construction (which is in the heart of our analysis) is relatively unknown to a broader scientific community, we included a very extensive appendix with a detailed account on this construction.

\setcounter{equation}{0}
\section{Spectral vs scaling dimensions \label{spectvsscal}}

Before defining the notion of a dimension for some generalized geometries, one has to answer the following question: What is the algebraic way of defining the dimension for the usual compact geometry? The answer is essentially given by  Weyl's theorem\\
\\
\textbf{Weyl's Theorem:}
\textit{Let $\Delta$ be the Laplace operator on a closed Riemannian manifold $\mathcal{M}$ of dimension $\mathrm{d}$. Let $N_\Delta (\omega)$ be the number of eigenvalues $\lambda_k$ of $\Delta$, counting multiplicities, less then $\omega$, i.e.\ $N_\Delta (\omega)$ is the counting function
\begin{eqnarray}\label{counting}
N_\Delta (\omega) := \#\{ \lambda_k (\Delta)\ :\ \lambda_k (\Delta)\leq \omega\}\ .
\end{eqnarray}
Then
\begin{eqnarray}\label{Weyl}
\lim_{\omega\rightarrow\infty}\frac{N_\Delta (\omega)}{\omega^{\frac{\mathrm{d}}{2}}}=\frac{Vol (\mathcal{M})}{(4\pi)^{\frac{\mathrm{d}}{2}}\Gamma(\frac{\mathrm{d}}{2}+1)}\ ,
\end{eqnarray}
where $Vol (\mathcal{M})$ is the total volume of the manifold $\mathcal{M}$.}\\
\\
This theorem can be used to \textit{calculate} the dimension $\mathrm{d}$ in the usual case: only when $\mathrm{d}$ coincides with the standard dimension, the limit in (\ref{Weyl}) will take a finite non-zero value.

As the first example let us calculate the (spectral) dimension of a flat commutative 2d torus $\mathbb{T}^2 = \mathbb{S}^1\times \mathbb{S}^1$ with two radii $r$ and $R$. The algebra of the continuous functions on $\mathbb{T}^2$ is generated by $u=\exp\frac{2\pi\ii x}{r}$ and $v=\exp\frac{2\pi\ii y}{R}$ (with $x,\ y$ being the usual coordinates along the torus cycles)
\begin{eqnarray}\label{algebraA}
\forall a\in \mathcal{A}\equiv \mathcal{C}^\infty (\mathbb{T}^2) \ ,\ a:= \sum\limits_{(l,m)\in\mathbb{Z}^2}\!\! a(l,m)\, u^l v^m \ ,
\end{eqnarray}
for some Schwartz function $a:\mathbb{Z}^2\rightarrow \mathbb{C}$. The standard derivations of this algebra are defined on the generators as
\begin{eqnarray}\label{derivcomm}
\left\{\begin{array}{c}
                                      \partial_1 u = \frac{2\pi\ii}{r} u \ ,\ \partial_1 v = 0 \\
                                      \partial_2 u = 0 \ ,\ \partial_2 v = \frac{2\pi\ii}{R} v
                                    \end{array}\right. \
\end{eqnarray}
and extended to the full algebra $\mathcal{A}$ by the Leibnitz rule.
Then the spectrum of its standard Laplacian, $\triangle_{com}$, is given by (we define the Laplacian with $\frac{1}{4\pi^2}$ factor\footnote{The origin of this factor is easily understood from (\ref{derivcomm}) defined with the factor of $2\pi$, which in its turn is the consequence of the definition of the generators $u$ and $v$ with $2\pi$ in the exponent.})
\begin{eqnarray}\label{spec_comm}
Spec(\triangle_{com}) = \left\{ \frac{n_1^2}{r^2} + \frac{n_2^2}{R^2} \ ,\ n_1,n_2 \in \mathbb{Z} \right\} \ .
\end{eqnarray}
The counting function is easily calculated
\begin{eqnarray}\label{counting_torus}
N_\Delta (\omega) \sim \int\limits_{\frac{n_1^2}{r^2} + \frac{n_2^2}{R^2} \leq \omega} d n_1 d n_2 = \pi r R \omega \ .
\end{eqnarray}
Using this in (\ref{Weyl}) we get
\begin{eqnarray}\label{dim_torus}
\lim_{\omega\rightarrow\infty}\frac{N_\Delta (\omega)}{\omega^{\frac{\mathrm{d}}{2}}} =
{ \left\{
\begin{array}{c}
\infty\ ,\ \mathrm{d}<2 \\
\pi r R\ ,\ \mathrm{d}=2 \\
0\ ,\ \mathrm{d}>2 \\
\end{array}
\right. } \ \ ,
\end{eqnarray}
so we conclude that in the case of a commutative torus the spectral dimension is $\mathrm{d}=2$ as it should be. Actually, using Weyl's theorem we can also recover the area of a 2d torus:
\begin{eqnarray}\label{area_torus}
\pi r R =  \frac{Area (\mathbb{T}^2)}{(4\pi)\Gamma(2)}\ \Rightarrow\ Area (\mathbb{T}^2)  = 4\pi^2 r R \ .
\end{eqnarray}
This gives (probably the most elaborate way of calculating) area or volume in general of any compact Riemanniean manifold.

As the second example of a less trivial application of the concept of a spectral dimension we would like to consider a noncommutative torus $\mathbb{T}^2_\theta$. We also use this example as an opportunity to introduce some definitions and notations that will be used later. By a noncommutative torus  we mean a $C^*$-algebra $\mathcal{A}_\theta$ generated by two unitary elements, $\mathbf{u}$ and $\mathbf{v}$ subject to the defining relation
\begin{eqnarray}\label{uvcommutator}
\mathbf{vu}=\exp(2\pi\ii\theta) \mathbf{uv},
\end{eqnarray}
where $\theta$ can be taken to belong to the interval $[0,1)$. The case of $\theta = 0$ corresponds to the commutative torus considered above. Then the arbitrary element of $\mathcal{A}_\theta$ will look exactly as in (\ref{algebraA}) (but now the order is important and some choice should be made)
\begin{eqnarray}\label{algebraAtheta}
\forall a\in \mathcal{A}_\theta \ ,\ a:= \sum\limits_{(l,m)\in\mathbb{Z}^2}\!\! a(l,m)\, \mathbf{u}^l \mathbf{v}^m \ ,
\end{eqnarray}
with exactly the same definition of the derivatives (for simplicity here we put $r=R=1$)
\begin{eqnarray}\label{derivnoncomm}
\left\{\begin{array}{c}
                                      \partial_1 \mathbf{u} = 2\pi\ii \mathbf{u} \ ,\ \partial_1 \mathbf{v} = 0 \\
                                      \partial_2 \mathbf{u} = 0 \ ,\ \partial_2 \mathbf{v} = 2\pi\ii \mathbf{v}
                                    \end{array}\right. \ ,
\end{eqnarray}
which again extended to the full algebra $\mathcal{A}_\theta$ by the Leibnitz rule.
Then it is obvious that the spectrum of the relevant (noncommutative) Laplace operator is the same as in (\ref{spec_comm}), which immediately leads to the conclusion that the spectral dimension of a noncommutative torus $\mathbb{T}^2_\theta$ is the same as in the commutative case, $\mathrm{d}=2$.

Now we would like to define the dimension for a somewhat more general geometry. Before taking on the noncommutative geometry of our interest, a fuzzy torus, we will illustrate our main idea by a two-dimensional toy model: a torus of one commutative and one fuzzy dimensions. First, we explain what we mean by one-dimensional fuzziness. We take as the definition of a torus with 1d fuzziness the geometry defined by the generalized Laplacian with the spectrum (\ref{spec_comm}) truncated along $R$-direction\footnote{Actually, as we mentioned above, to define a geometry one needs the full spectral triple with a Dirac operator, an algebra and a Hilbert space. Using the same algebra this geometry does not correspond to a proper spectral triple because there is no corresponding Dirac operator \emph{with compact resolvent}. Fortunately, for our demonstrative purposes this Laplacian will suffice.}\label{footnote1}
\begin{eqnarray}\label{spec_nc}
Spec(\triangle_{1fuzzy}) = \left\{ \frac{n_1^2}{r^2} + \frac{n_2^2}{R^2} \ ,\ n_1,n_2 \in \mathbb{Z}\ ,\ |n_2| \leq N \right\} \ .
\end{eqnarray}

How would one define the dimension of such geometry? As it was argued in \cite{Pinzul:2010ct} the spectral dimension defined via the obvious generalization of Weyl's theorem seems to be the most natural definition of a \textit{physical} dimension. We would like to generalize Weyl's theorem in such a way that it could be used to define \textit{an effective}, or \textit{scaling}, dimension. From (\ref{Weyl}) it is clear that the spectral dimension defined by Weyl's theorem is an UV-dimension, i.e.\ the dimension as seen in an experiment that can probe any scale. Obviously this is not the case in reality. \textit{Define} the scaling dimension as\footnote{In this definition we assume that the cut-off dependence of volume in Weyl's theorem is not important. In \cite{Gregory:2012an}, we showed that by somewhat reversing the arguments one can use the generalization of Weyl's theorem to calculate the quantum corrections to area, which was demonstrated on the example of a fuzzy sphere.} \cite{Pinzul:2010ct}
\begin{eqnarray}\label{scaling_dim}
d(\omega) := 2 \frac{d \ln N_\Delta (\omega)}{d \ln \omega}\ .
\end{eqnarray}
This defines the dimension seen in an experiment that can probe the physics only up to the scale $\omega$. The scale is defined in terms of the spectrum of a relevant physical Laplacian, i.e.\ an operator that controls the dynamics 
used to probe the geometry of the space-time.\footnote{Recall, that the typical coupling of the matter sector to the geometry has a form of the Dirac action, $S_{mat}\sim \langle\psi | \mathcal{D} |\psi \rangle$. \cite{Connes:1995tu}} Obviously the definition (\ref{scaling_dim}) makes sense only if the cut-off scale $\omega$ is large enough, so the dependence of $N_\Delta (\omega)$ on $\omega$ could be considered as smooth (below more on this). The difference between the UV-dimension and the scaling one could be readily seen in any matrix geometry, i.e.\ when the relevant operators have finite spectra. In this case the counting function goes to a constant when the cut-off $\omega$ goes to infinity. This means that any matrix geometry has a UV-dimension equal to zero. At the same time, it seems very natural that if the spectrum is truncated at very high energy, we will not be able to tell the smooth geometry from the matrix one. Hence in any accessible experiment we will see the matrix geometry as a smooth one with some defined dimension (and probably with some quantum corrections). This observation makes the concept of a scaling dimension a very natural one.

\begin{figure}[htb]
\begin{center}
\leavevmode
\includegraphics[scale=0.4]{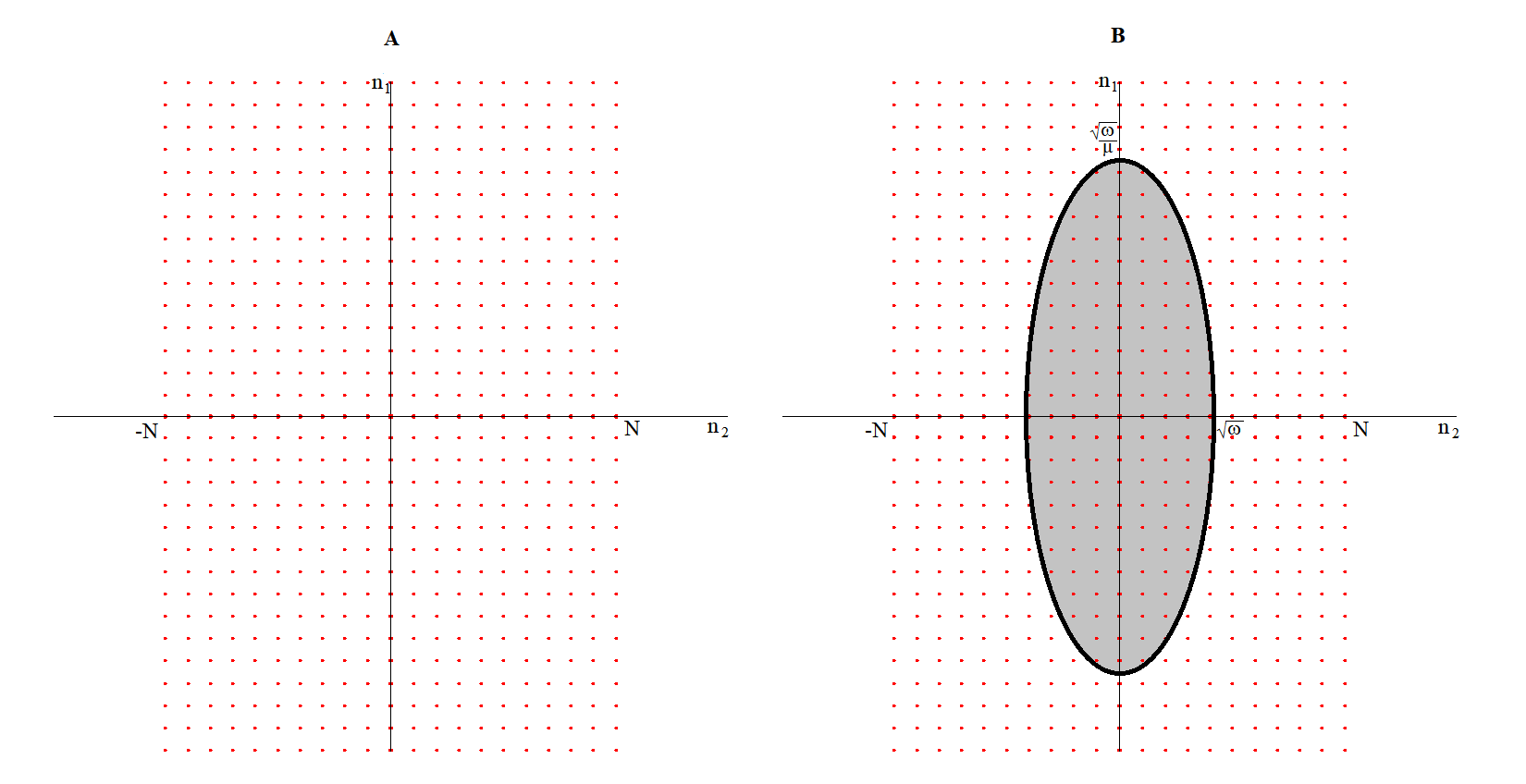}
\end{center}
\caption{\sl A. The structure of a typical spectrum with the $n_2$-direction truncated at $N$; B. The solid curve $\mu^2 n_1^2 + n_2^2 = \omega$ represents a cut-off (we set $R = 1$). All the points of the spectrum inside the shadowed area are below the cut-off.}
\label{fig:spec}
\end{figure}

We now will apply this concept to the geometry defined by (\ref{spec_nc}). We will analyze this somewhat not-well defined case (see the footnote on p.\pageref{footnote1}) in some details because it demonstrates the great variety of the physical situations depending on the scale and also because the analysis of the case when $r\sim R$ appears to be technically very similar to our main model - a fuzzy torus.  We will see that the situation (i.e.\ the interpretation of the outcome of the ``experiment'') drastically depends on the interplay between the two parameters: the aspect ratio $\mu:= \frac{R}{r}$ and the ``scale of fuzziness'' $N$. This becomes evident if one re-writes the spectrum (\ref{spec_nc}) as
\begin{eqnarray}\label{spec_nc1}
Spec(\triangle_{1fuzzy}) = \left\{ \frac{1}{R^2} \left(\mu^2 n_1^2 + n_2^2 \right)\ ,\ n_1,n_2 \in \mathbb{Z}\ ,\ |n_2| \leq N \right\} \ .
\end{eqnarray}
The structure of a typical spectrum can be represented graphically as on Fig.\ref{fig:spec}A, while Fig.\ref{fig:spec}B gives the graphical answer for the counting function (\ref{counting}).

\smallskip
\noindent\textbf{The case of $\mu\gg 1$ (large fuzzy dimension)}

\noindent i) We will always assume that $N$ is finite but large, $N\gg 1$. At the beginning, we assume that $N$ is large enough, so the following inequalities hold
\begin{eqnarray}\label{mu_large}
1\ll \omega R^2 < \mu^2 \ \mathrm{and\ at\ the\ same\ time}\ \omega R^2 < N^2\ .
\end{eqnarray}
From the point of view of Fig.\ref{fig:spec}B these conditions mean that the $n_1$ semi-axis of the cut-off ellipse is so small that no state with $n_1 \neq 0$ will contribute to the counting function. At the same time the number of states with non-zero $n_2$ is large enough to allow the application of the formula (\ref{scaling_dim}) for the scaling dimension. Calculating the counting function, we get
\begin{eqnarray}\label{scaling_dim1}
N_\Delta (\omega) \sim 2 \sqrt{\omega} R\ \Rightarrow\
d(\omega) = 2 \frac{d \ln N_\Delta (\omega)}{d \ln \omega} = 1\ .
\end{eqnarray}
So we arrive at a very natural and expected result: if the experiment probes scales below the energy needed to excite the first Kaluza-Klein (KK) mode it does not see the corresponding compactified dimension. (And at the same time, the other, fuzzy, dimension looks perfectly commutative!)

\noindent ii) Upon increasing the cut-off scale $\omega$ the states with $n_1 \neq 0$ will start contributing to the counting function. But only when many of them will enter, i.e.\ when $\omega R^2 \gg \mu^2$, (so one can pass from the sum to the integral as, e.g., in (\ref{counting_torus})) one can start using (\ref{scaling_dim}) to determine the dimension. This can happen either when a) $\omega R^2$ is still less then $N^2$ or b) $\omega R^2 > N^2$ (but still of the order of $N$) or c) $\omega R^2 \gg N^2$. Let us analyze these possibilities separately.

\noindent a) $\omega R^2 < N^2$ means that now we are effectively probing the geometry of a torus $\mathbb{T}^2$, so the counting function and the scaling dimension are given by (\ref{counting_torus}, \ref{dim_torus})
\begin{eqnarray}\label{scaling_dim2a}
N_\Delta (\omega) \sim \pi \omega r R\ \Rightarrow\
d(\omega) = 2 \frac{d \ln N_\Delta (\omega)}{d \ln \omega} = 2\ .
\end{eqnarray}
Increasing further the cut-off $\omega$, we will be in the situation (b) below.

\noindent b) $\omega R^2 > N^2$ corresponds to the situation when $n_2$ semi-axis, see Fig.\ref{fig:spec}B, is greater then the truncation $N$. In this case one can easily calculate the counting function
\begin{eqnarray}\label{counting2b}
N_\Delta (\omega) &\sim & 4\int\limits_0^N d n_2 \int\limits_0^{\frac{\sqrt{\omega}R}{\mu}\sqrt{1-\frac{n_2^2}{\omega R^2}}} d n_1 = \frac{4\omega R^2}{\mu}\int\limits_0^{\frac{N}{\sqrt{\omega}R}}d x \sqrt{1-x^2} = \nonumber\\
&=& \frac{2\omega R^2}{\mu}\left( \frac{N}{\sqrt{\omega}R}\sqrt{1 - \frac{N^2}{\omega R^2}} + \arcsin \left( \frac{N}{\sqrt{\omega}R} \right)\right)\ .
\end{eqnarray}
If we formally take $\omega R^2 = N^2$ we will get back the commutative result (\ref{counting_torus}) or (\ref{scaling_dim2a}), as it should be, because the experiment still would not know anything about the truncation. Using (\ref{counting2b}) in the definition of the scaling dimension (\ref{scaling_dim}) we get
\begin{eqnarray}\label{scaling_dim2b}
d(\omega) = 2 \left( 1+\frac{N}{\sqrt{\omega} R} \frac{\sqrt{1-\frac{N^2}{\omega R^2}}}{\arcsin \left( \frac{N}{\sqrt{\omega}R} \right)} \right)^{-1} \ .
\end{eqnarray}
It is clear that this expression describes a flow from $d(\omega) = 2$ when $\omega R^2 \rightarrow N^2$ to $d(\omega) = 1$ when $\omega R^2 \rightarrow \infty$.

\noindent c) If the experiment will see the large number of $n_1$-states only when $\omega R^2 \gg N^2$, the counting function will be constant (or almost constant) for $\omega R^2 > N^2$ (up to same characteristic scale $\omega_0$ for which one can say that there are ``many'' $n_1$-states, i.e.\ $\sqrt{\omega_0}R \gg \mu$). Then applying our definition of the scaling dimension (\ref{scaling_dim}) we get that in this region $d(\omega) = 0$.

When $\omega \geq \omega_0$ we again can use (\ref{scaling_dim}) to obtain $d(\omega)$. The calculation is the same as in (\ref{counting2b}) but now $\omega R^2 \gg N^2$ so one gets $d(\omega) = 1$.

\noindent iii) Now we would like to discuss the transitional regime. In the case under study this corresponds to $\omega R^2 > \mu^2$, but $\frac{\sqrt{\omega}R}{\mu}$ is still of order of one (not too large).

This is the standard Kaluza-Klein situation. When $\omega R^2 > \mu^2$, we will start exiting one by one the KK modes. While $\omega R^2$ is still not too big, we would continue to interpret this in the usual way but at some point the alternative interpretation - the emergence of a new dimension - might become more appropriate. The conclusive interpretation could be made only by using the experiments with better resolutions. In general, the interpretation of the transitional regimes is very subjective and could be treated either as the change in the geometry (dimension) or in the dynamics (degrees of freedom) of the model.

\begin{figure}[htb]
\begin{center}
\leavevmode
\includegraphics[scale=0.37]{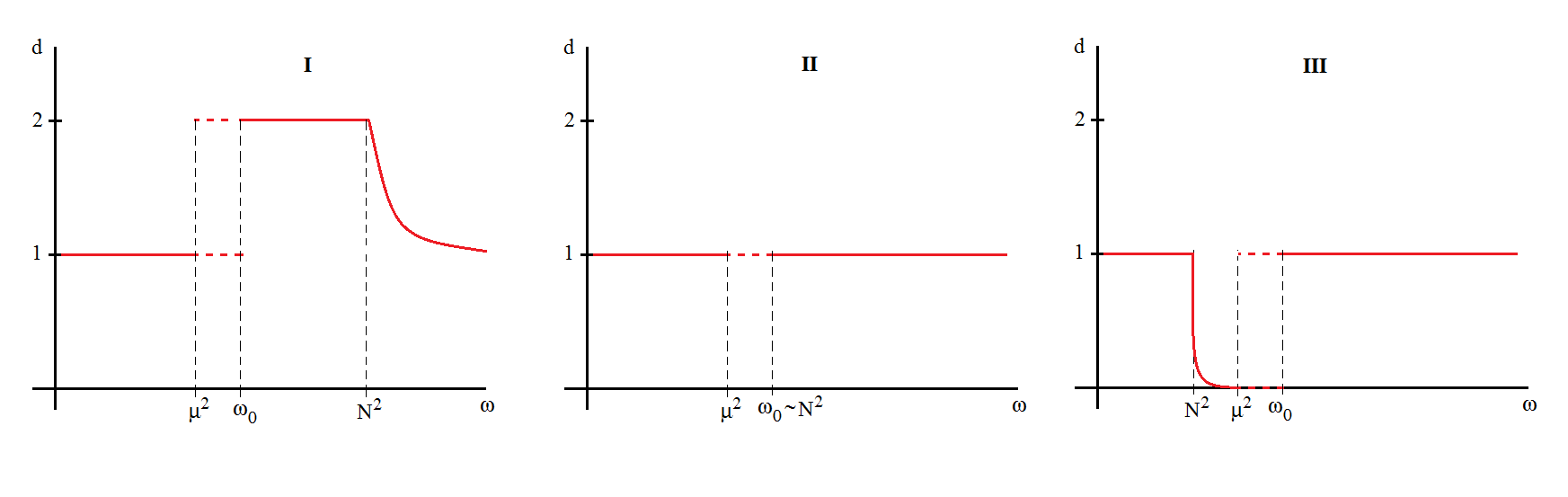}
\end{center}
\caption{\sl The typical behaviour of the scaling dimension for the case a large fuzzy dimension ($R$ is set to $1$). The figures I, II and III correspond to the cases considered in the point (ii): I is the combined (a)-(b) regime, while II and III correspond to the different possibilities for $\omega_0$ scale from the case (c). By a dashed line we denoted the transitional regime, see (iii), where the physical dimension is subject to interpretation.}
\label{fig:specdim1}
\end{figure}

Schematically, the behavior of the scaling spectral dimension for the case when $\mu\gg 1$ is shown on Fig.\ref{fig:specdim1}.

\smallskip
\noindent\textbf{The case of $\mu\sim 1$}

The case of $\mu\gg 1$ that we considered above is a very nice demonstration of the variety of different non-trivial regimes as well as their physical (``experimental'') interpretation. Here we will consider another important case when the fuzzy and commutative radii are about the same size. This case is very relevant for our further consideration of a more realistic model of a fuzzy torus. From the point of view of figure (\ref{fig:spec}), the case of $\mu\sim 1$ corresponds to having roughly a circle for the cut-off region. This immediately shows that we essentially have just two regimes: a) $1\ll \omega R^2 < N^2 $ and b) $N^2 < \omega R^2 $.

a) Because $1\ll \omega R^2 < N^2 $, we see that for the case of $\mu\sim 1$ the experiment will probe the part of the spectrum (\ref{spec_nc1}) that is exactly the same as in the commutative case (\ref{spec_comm}). So repeating the same arguments as in the case of $\mu\gg 1$ (case ($\mathrm{ii}_a$)) we immediately conclude that the spectral dimension as seen by the experiment will be $d(\omega)=2$ (cf. (\ref{scaling_dim2a})).

b) The case when $N^2 < \omega R^2 $ was considered above as well, and the resulting scaling dimension was given by (\ref{scaling_dim2b}). The observed dimension smoothly goes from 2 to 1.

\begin{figure}[htb]
\begin{center}
\leavevmode
\includegraphics[scale=0.37]{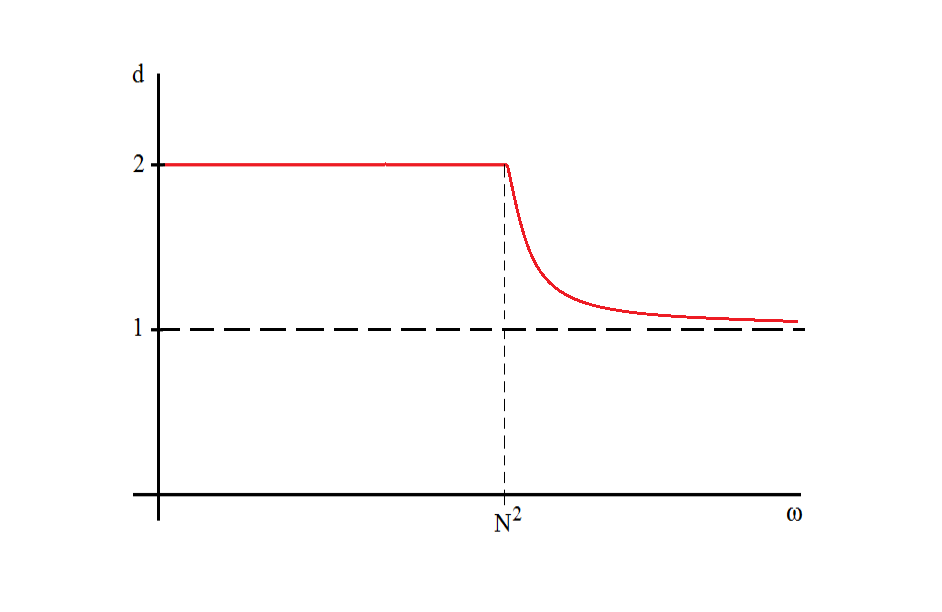}
\end{center}
\caption{\sl The typical behaviour of the scaling dimension for the case when the fuzzy and the commutative radii are the same (and set to $1$). }
\label{fig:specdim2}
\end{figure}

The only difference between the $\mu\gg 1$ and $\mu\sim 1$ cases is that the former one has an additional large parameter, $\mu$, that effectively introduces one more scale compared to the latter case. This explains the variety of possibilities in the situation with $\mu\gg 1$. The $\mu\sim 1$ case is schematically summarized in Fig.\ref{fig:specdim2}. 
In principle, there is one more situation: $\mu\ll 1$. It can be analysed in the complete analogy with the ones we have considered. We will not describe it as it does not contribute anything to the understanding of the idea.

\setcounter{equation}{0}
\section{Matrix Approximations to the Noncommutative Torus \label{se:approximation}}

Our construction is based on approximating a torus by some sort of a \emph{fuzzy} torus. It seems very natural to assume that the algebra $\mathcal{A}_\theta$ is recovered as some inductive limit of the matrix algebras $\mathrm{Mat}_{q}(\mathbb{C})$. This is equivalent to saying that $\mathcal{A}_\theta$ is an approximately finite (AF) algebra. It is well known that unfortunately this is not possible. The most straightforward way to see this is by looking at the \textit{K}-theory of $\mathcal{A}_\theta$ and of any AF-algebra. While $K_n(\mathbb{T}^2_\theta)=K_n(\mathbb{T}^2)\equiv \mathbb{Z}\oplus\mathbb{Z}$, $n=0,1$, $K_1$ of any AF-algebra is trivial \cite{WeggeOlsen:1994}. In \cite{Pimsner:1980,Landi:1999ey} it was clarified how one should construct and interpret the finite matrix approximation of the algebra of a noncommutative torus for an arbitrary $\theta$. Because we will not use this construction in this work, we refer to \cite{Landi:1999ey} for all the details and to the review~\cite{Lizzi:1997yr} for the broader context.

There is however a construction, due to Elliot and Evans (EE)~\cite{Elliott:1993} which shows that the algebra of the noncommutative torus is the inductive limit of the algebra of two copies of the algebra of matrices whose entries are functions on a circle $\mathbb T\equiv \mathbb{S}^1$. Due to the presence of these functions, the algebra is not approximatively finite, and since the $K$-theory of a circle is $\mathbb Z$ there is the required matching of $K$-theories. Note however that at the finite level the algebra corresponds to a topological sum (not a product!) of two circles, i.e. a one-dimensional space.

In this section we present only a short summary, while in the Appendix we give a more detailed review of the EE construction.

For the case of $\theta=\frac pq$ rational there is a finite dimensional representation of the relation~\eqref{uvcommutator} by two matrices called \emph{clock} and \emph{shift}\footnote{For coherence with the EE construction and their notations, our definition for these matrices is slightly different from the one which usually appears in the physics literature.}:
\begin{eqnarray}\label{clockshift}
u_q := \left(
                   \begin{array}{ccccc}
                     1 & 0 & 0 & \cdots & 0 \\
                     0 & \xi^{-1} & 0 & \cdots & 0 \\
                     0 & 0 & \xi^{-2} & \cdots & 0 \\
                     \vdots & \vdots & \vdots & \ddots & \vdots \\
                     0 & 0 & 0 & 0 & \xi^{1-q} \\
                   \end{array}
                 \right) \ ,\
v_q (z):= \left(
                   \begin{array}{cccccc}
                     0 & 0 & 0 & \cdots & 0 & 1 \\
                     1 & 0 & 0 & \cdots & 0 & 0 \\
                     0 & 1 & 0 & \cdots & 0 & 0 \\
                     \vdots & \vdots & \vdots & \ddots & 0 & 0 \\
                     0 & 0 & 0 & \cdots & 1 & 0 \\
                   \end{array}
                 \right) \ ,\ \xi = \ee^{2\pi\ii\frac{p}{q}}\ .
\end{eqnarray}
The algebra generated by these two matrices is usually called a \emph{fuzzy torus}, and is of course $\mathrm{Mat}_q(\mathbb C)$. As we said no limit of this algebra could reproduce a noncommutative torus.

The EE construction starts by considering two (sequences of) rational numbers
\begin{equation}
\frac pq<  \theta < \frac{p'}{q'}
\end{equation}
such that in the limit $q,q'\to\infty$ they both converge to $\theta$,  two sets of projections $P_{ii},P'_{i'i'}$ ($i=1\ldots q,\, i'=1\ldots q')$ and two sets of partial isometries $P_{ij},P'_{i'j'}$. Projections and isometries are elements of the algebra of a noncommutative torus, $\mathcal A_\theta$. They behave like matrix units, i.e.\ satisfy the relations
\begin{equation}
P_{ij}P_{kl}=\delta_{jk}P_{il}
\end{equation}
 and are obtained one from another by the action of some translation isomorphism $\alpha$, e.g., $\alpha^{i-1}(P_{11})=P_{ii}$ and the analogous formulas for primed quantities and isometries. Except that $\alpha^{q-1}(P_{21}) = z P_{1q}$ where $z\in\mathcal A_\theta$ is an unitary element of the algebra. The construction runs parallel in the primed and unprimed sectors enabling the building of two ``towers'' of elements of the algebra. The subalgebra of $\mathcal{A}_\theta$ generated by these towers is isomorphic to $\mathrm{Mat}_q (\mathcal{C}^\infty (\mathbb{S}^1)) \oplus \mathrm{Mat}_q (\mathcal{C}^\infty (\mathbb{S}^1))$ and it has two unitaries which generalise the clock and shift matrices above (see (\ref{finalapprox})):
\begin{eqnarray}\label{finalapprox1}
\mathbf{U} = \mathcal{C}_q \oplus \mathcal{S}_{q'} (z') \equiv \left(
                                                             \begin{array}{cc}
                                                               \mathcal{C}_q & 0_{q\times q'} \\
                                                               0_{q'\times q} & \mathcal{S}_{q'} (z') \\
                                                             \end{array}
                                                           \right) \ , \nonumber\\
\mathbf{V} = \mathcal{S}_q (z) \oplus \mathcal{\bar{C}}_{q'}  \equiv \left(
                                                             \begin{array}{cc}
                                                               \mathcal{S}_q (z)& 0_{q\times q'} \\
                                                               0_{q'\times q} & \mathcal{\bar{C}}_{q'} \\
                                                             \end{array}
                                                           \right)    \ ,
\end{eqnarray}
where now
\begin{eqnarray}\label{CSmatrix}
\mathcal{C}_q := \left(
                   \begin{array}{ccccc}
                     1 & 0 & 0 & \cdots & 0 \\
                     0 & \xi^{-1} & 0 & \cdots & 0 \\
                     0 & 0 & \xi^{-2} & \cdots & 0 \\
                     \vdots & \vdots & \vdots & \ddots & \vdots \\
                     0 & 0 & 0 & 0 & \xi^{1-q} \\
                   \end{array}
                 \right) \ ,\
\mathcal{S}_q (z):= \left(
                   \begin{array}{cccccc}
                     0 & 0 & 0 & \cdots & 0 & z \\
                     1 & 0 & 0 & \cdots & 0 & 0 \\
                     0 & 1 & 0 & \cdots & 0 & 0 \\
                     \vdots & \vdots & \vdots & \ddots & 0 & 0 \\
                     0 & 0 & 0 & \cdots & 1 & 0 \\
                   \end{array}
                 \right) \ ,\ \xi = \ee^{2\pi\ii\frac{p}{q}}\ .
\end{eqnarray}

The unitaries $\mathbf U$ and $\mathbf V$ provide a good (and even, in some sense, the best) approximation in norm of $\mathbf u$ and $\mathbf v$ (\ref{uvcommutator}) as the matrices become larger and larger, i.e. $q,q'\to\infty$ (see the details in the appendix). They generate the algebra $\mathrm{Mat}_q (\mathcal{C}^\infty (\mathbb{S}^1)) \oplus \mathrm{Mat}_q (\mathcal{C}^\infty (\mathbb{S}^1))$ (two summands corresponding to two towers), whose inductive limit is~$\mathcal A_\theta$.

\setcounter{equation}{0}
\section{Truncation map \label{se:truncation}}

To continue our discussion, we need to specify the nature of the integers $q,q'$ appearing in the EE construction. The only condition on $q,q'$ and $p,p'$ (or $\beta , \beta'$) is given in (\ref{qqpp}), i.e.\ that $\left(
                          \begin{array}{cc}
                            p' & p \\
                            q' & q \\
                          \end{array}
                        \right) \in SL(2,\mathbb{Z}) $. This happens to be exactly the condition satisfied by two consecutive approximations of $\theta$ by continued fractions, see e.g. \cite{Khinchin:1964}:
\begin{eqnarray}\label{fractions}
\frac{p_{2n}}{q_{2n}} < \theta < \frac{p_{2n-1}}{q_{2n-1}} \ ,
\end{eqnarray}
see \cite{Landi:2004sc} for the details. From now on, we take $q=q_{2n}$ and $q' = q_{2n-1}$ and denote $\mathcal{A}_n :=\mathcal{A}_{qq'}$, $P_n := P$ and so on.

Because we are interested in the finite dimensional approximations of $\mathbb{T}^2_\theta$ the first question to answer is how any particular element of the algebra $\mathcal{A}_\theta$ is approximated by an element of $\mathcal{A}_n$. Towards this end let us define the ``truncation'' map $\Gamma_n$: $\mathcal{A}_\theta \rightarrow \mathcal{A}_n$.
\begin{eqnarray}\label{truncation}
\forall a\in \mathcal{A}_\theta\ ,\ \Gamma_n (a):= \sum\limits_{(l,m)\in\mathbb{Z}^2}\!\! a(l,m)\, \mathrm{\mathbf{U}}_n^l \mathrm{\mathbf{V}}_n^m \ ,
\end{eqnarray}
where by $\mathrm{\mathbf{U}}_n$ and $\mathrm{\mathbf{V}}_n$ we denoted the generators (\ref{finalapprox1}) for the case when $q=q_{2n}$ and $q'=q_{2n-1}$.

Using $(\mathcal{C}_q )^q = \mathds{1}_q\ ,\  (\mathcal{S}_q (z))^q = z\mathds{1}_q$ (and the same for $q'$) we have:
\begin{eqnarray}
&&\forall r\in\mathbb{Z}\ \ \exists l\in\mathbb{Z}\ \mathrm{and}\ k=\overline{0,q-1}\ :\ r=lq+k\ \Rightarrow\ (\mathcal{S}_{q}(z))^r = z^l (\mathcal{S}_{q}(z))^k \ ,\\
&&\forall m\in\mathbb{Z}\ \ \exists s\in\mathbb{Z}\ \mathrm{and}\ i=\overline{0,q-1}\ :\ m=sq+i-\left[\frac{q}{2}\right]\ \Rightarrow\ (\mathcal{C}_{q})^m = (\mathcal{C}_{q})^{i-\left[\frac{q}{2}\right]} \ ,
\end{eqnarray}
where by $[\cdots]$, as usual, we denote the integer part.
Then we can re-write the first entry of the direct sum $\Gamma_n (a)=:\mathbf{a}^{(n)}(z)\oplus\mathbf{a}'^{(n)}(z')$ in the following form
\begin{eqnarray}
\mathbf{a}^{(n)}(z)=\sum\limits_{(m,r)\in\mathbb{Z}^2}\!\! a(m,r)\, (\mathcal{C}_{q})^m (\mathcal{S}_{q}(z))^r = \\
=\sum\limits^{q_{2n}-1}_{i=0}\sum\limits_{s\in\mathbb{Z}}\sum\limits^{q_{2n}-1}_{k=0}\sum\limits_{l\in\mathbb{Z}}\!\! a(sq_{2n}+i-\left[\frac{q_{2n}}{2}\right],lq_{2n}+k) z^l (\mathcal{C}_{q_{2n}})^{i-\left[\frac{q_{2n}}{2}\right]} (\mathcal{S}_{q_{2n}}(z))^k = \\
=\sum\limits^{q_{2n}-1}_{i,k=0}\sum\limits_{l\in\mathbb{Z}}\left( \sum\limits_{s\in\mathbb{Z}}\!\! a(sq_{2n}+i-\left[\frac{q_{2n}}{2}\right],lq_{2n}+k) \right) z^l (\mathcal{C}_{q_{2n}}(z))^{i-\left[\frac{q_{2n}}{2}\right]} (\mathcal{S}_{q_{2n}}(z))^k =: \\
=: \sum\limits^{q_{2n}-1}_{i,k=0}\sum\limits_{l\in\mathbb{Z}} a^{(n)}(i,k;l) z^l (\mathcal{C}_{q_{2n}})^{i-\left[\frac{q_{2n}}{2}\right]} (\mathcal{S}_{q_{2n}}(z))^k \ ,
\end{eqnarray}
where
\begin{eqnarray}\label{an}
a^{(n)}(i,k;l) := \sum\limits_{s\in\mathbb{Z}}\!\! a(sq_{2n}+i-\left[\frac{q_{2n}}{2}\right],lq_{2n}+k) \ .
\end{eqnarray}
Then for $\Gamma_n (a)$ we finally have
\begin{eqnarray}\label{truncation1}
\Gamma_n (a)= \left(\sum\limits_{m,r=0}^{q_{2n}-1}\sum\limits_{l\in\mathbb{Z}} a^{(n)}(m,r;l) z^l (\mathcal{C}_{q_{2n}})^{m-\left[\frac{q_{2n}}{2}\right]} (\mathcal{S}_{q_{2n}}(z))^r \right) \oplus \nonumber \\
\oplus\left(\sum\limits_{m',r'=0}^{q_{2n-1}-1}\sum\limits_{l'\in\mathbb{Z}} a'^{(n)}(m',r';l') z'^{l'} (\mathcal{S}_{q_{2n-1}}(z'))^{m'} (\mathcal{\bar{C}}_{q_{2n-1}})^{r'-\left[\frac{q_{2n-1}}{2}\right]} \right)\ .
\end{eqnarray}
The result (\ref{truncation1}) for the truncation map is slightly different from the one found in \cite{Landi:2004sc}, but equivalent. The more symmetric form (compared to the one in \cite{Landi:2004sc}) will be useful in the construction of the modified derivatives, to which we now proceed.

\setcounter{equation}{0}
\section{Deformed derivatives and spectrum}\label{se:derivatives}
In this section we show that the noncommutative (topological) space corresponding to $\mathcal{A}_n$, equipped with two deformed (and approximate) derivatives, describes the approximation to the geometry of the noncommutative torus $\mathbb{T}^2_\theta$. We also find the spectrum of these derivatives as the first step towards the spectral dimension, which we discuss in the next section.

\subsection{Derivatives}
The natural condition on the derivations in $\mathcal{A}_n$ would be that they leave the eigen-spaces of the projectors $P_n$ and $P'_n$ invariant, i.e.\ that the block-diagonal structure (\ref{finalapprox1}) is preserved under the derivation. One would like to have some consistent truncation or deformation of the standard derivatives on $\mathcal{A}_\theta$ defined in (\ref{derivnoncomm})
\begin{eqnarray}\label{deriv}
\forall a\in \mathcal{A}_\theta\ ,\ \left\{\begin{array}{c}
                                      \partial_1 \mathbf{u} = 2\pi\ii \mathbf{u} \ ,\ \partial_1 \mathbf{v} = 0 \\
                                      \partial_2 \mathbf{u} = 0 \ ,\ \partial_2 \mathbf{v} = 2\pi\ii \mathbf{v}
                                    \end{array}\right.
                                    \Leftrightarrow
                                    \left\{\begin{array}{c}
                                      \partial_1 a = 2\pi\ii \sum\limits_{(l,m)\in\mathbb{Z}^2}\!\! l\, a(l,m)\, \mathbf{u}^l \mathbf{v}^m \\
                                      \partial_2 a = 2\pi\ii \sum\limits_{(l,m)\in\mathbb{Z}^2}\!\! m\, a(l,m)\, \mathbf{u}^l \mathbf{v}^m
                                    \end{array}\right. \ .
\end{eqnarray}

While (\ref{deriv}) defines the derivatives consistent with the Leibnitz rule, unfortunately it is impossible to define derivatives of $\mathcal{A}_n$ with the same property and at the same time respecting the block-diagonal structure.\footnote{It is also true in the zero-dimensional approximation described in \cite{Landi:1999ey}. The reason is that (\ref{deriv}) is incompatible with, e.g. $(\mathcal{C}_q )^q = \mathds{1}_q$.} Instead we define the approximate derivatives, $\nabla_i$, $i=1,2$, using as the motivation the image of the derivatives (\ref{deriv}) under the truncation map (\ref{truncation})
\begin{eqnarray}\label{deriv_deformed}
\nabla_i \Gamma_n (a):= \Gamma_n (\partial_i a) + \mathcal{O}(\cdots) \ ,
\end{eqnarray}
where by $\mathcal{O}(\cdots)$ we denote the terms that vanish in the $q,q' \rightarrow \infty$ limit. The choice of these terms is made in such a way as to insure that the action of $\nabla_i$ is diagonal in the representation (\ref{truncation1}). Using (\ref{truncation1}) and (\ref{deriv}) we arrive at the explicit expressions:
\begin{eqnarray}\label{deriv_devormed1}
&&\nabla_1\Gamma_n (a):= 2\pi\ii\left(\sum\limits_{m,r=0}^{q_{2n}-1}\sum\limits_{l\in\mathbb{Z}} \left( m - \left[\frac{q_{2n}}{2}\right]\right)\,a^{(n)}(m,r;l) z^l (\mathcal{C}_{q_{2n}})^{m-\left[\frac{q_{2n}}{2}\right]} (\mathcal{S}_{q_{2n}}(z))^r \right. \oplus \nonumber \\
&\oplus&\left.\sum\limits_{m',r'=0}^{q_{2n-1}-1}\sum\limits_{l'\in\mathbb{Z}} (l' q_{2n-1}+m')\,a'^{(n)}(m',r';l') z'^{l'} (\mathcal{S}_{q_{2n-1}}(z'))^{m'} (\mathcal{\bar{C}}_{q_{2n-1}})^{r'-\left[\frac{q_{2n-1}}{2}\right]} \right) \ , \nonumber \\
&&\nabla_2\Gamma_n (a):= 2\pi\ii\left(\sum\limits_{m,r=0}^{q_{2n}-1}\sum\limits_{l\in\mathbb{Z}} \left( l q_{2n}+r \right)\,a^{(n)}(m,r;l) z^l (\mathcal{C}_{q_{2n}})^{m-\left[\frac{q_{2n}}{2}\right]} (\mathcal{S}_{q_{2n}}(z))^r \right. \oplus \nonumber \\
&\oplus&\left.\sum\limits_{m',r'=0}^{q_{2n-1}-1}\sum\limits_{l'\in\mathbb{Z}} \left( r'-\left[\frac{q_{2n-1}}{2}\right] \right)\,a'^{(n)}(m',r';l') z'^{l'} (\mathcal{S}_{q_{2n-1}}(z'))^{m'} (\mathcal{\bar{C}}_{q_{2n-1}})^{r'-\left[\frac{q_{2n-1}}{2}\right]} \right) \ .
\end{eqnarray}
The $\nabla_i$ are only \emph{approximate} derivatives because they satisfy the Leibnitz rule only in the limit of large  $n$.

\subsection{Spectrum}

Ultimately we are interested in the spectrum of the deformed analog of the Laplacian, which we define in the complete analogy with the commutative case (we do not include the factor of $\frac{1}{4\pi^2}$ as in the commutative case because it does not affect the spectral dimension):
\begin{eqnarray}\label{deformed_Laplace}
\Delta_{(n)} = -\nabla_1^2 - \nabla_2^2\ .
\end{eqnarray}
We consider the integer $n$ (and therefore $q_{2n}$, $q_{2n+1}$ etc.) to indicate a physical cutoff. In other words the deformed Laplacian represents the geometry we want to investigate.

Let us begin with the spectrum of the operator $\nabla_1$. In the non-truncated case the spectrum is well-known, see (\ref{deriv})
\begin{eqnarray}\label{specnabla1}
Spec (\nabla_1) = 2\pi\ii \mathbb{Z} \ .
\end{eqnarray}

Using the definitions (\ref{Cq}) and (\ref{Sq}) of $C_q$ and $S_q (z)$ in terms of the matrix units one can easily prove the orthogonality relations
\begin{eqnarray}\label{orthogonal}
\mathrm{Tr}\left[(\mathcal{S}_{q}(z)^\dagger)^l(\mathcal{C}_{q}^\dagger)^p(\mathcal{C}_{q})^m (\mathcal{S}_{q}(z))^r \right] = q\beta_q\delta_{lr}\delta_{pm} \ .
\end{eqnarray}
The appearance of the constant $\beta_q$ is due to the normalization. In terms of the continuous fraction expansion $\beta_q$ has the following form:
\begin{eqnarray}\label{beta}
\beta_q = \left\{ \begin{array}{ll}
                    q_{2n-1}(\theta_{2n-1} - \theta)\ , & q=q_{2n} \\
                    q_{2n}(\theta - \theta_{2n})\ , & q=q_{2n-1}
                  \end{array}
 \right. \ .
\end{eqnarray}
The following relation: $q_{2n}\beta_{2n}+q_{2n-1}\beta_{2n-1}=1$ holds. This insures that $\mathrm{Tr}\Gamma_n(\mathds{1})=1$.

Combining this result with the obvious fact that $C_q$ and $S_q (z)$ generate $\mathrm{Mat}_{q}(\mathcal{C^\infty}(\mathbb{S}^1))$ (and the same for the primed side of the construction),
we can choose the orthogonal basis for $\mathcal{A}_n \cong \mathrm{Mat}_{q_{2n}}(\mathcal{C^\infty}(\mathbb{S}^1))\oplus \mathrm{Mat}_{q_{2n-1}}(\mathcal{C^\infty}(\mathbb{S}^1))$ as
\begin{eqnarray}\label{basis}
\left\{ z^l (\mathcal{C}_{q_{2n}})^{m-\left[\frac{q_{2n}}{2}\right]} (\mathcal{S}_{q_{2n}}(z))^r \oplus 0_{q_{2n-1}} \right\}\bigcup
\left\{ 0_{q_{2n}} \oplus z'^{l'} (\mathcal{S}_{q_{2n-1}}(z'))^{m'} (\mathcal{\bar{C}}_{q_{2n-1}})^{r'-\left[\frac{q_{2n-1}}{2}\right]} \right\} \ ,
\end{eqnarray}
where $0_{q}$ is a zero $q\times q$ matrix and all indices as in (\ref{truncation1}) or (\ref{deriv_devormed1}). From (\ref{deriv_devormed1}) it is immediately obvious that this basis is made of the eigen-vectors of $\nabla_1$
\begin{eqnarray}
&&\nabla_1 \left( z^l (\mathcal{C}_{q_{2n}})^{m-\left[\frac{q_{2n}}{2}\right]} (\mathcal{S}_{q_{2n}}(z))^r \oplus 0_{q_{2n-1}} \right) = \nonumber\\
&&=2\pi\ii \left( m-\left[\frac{q_{2n}}{2}\right] \right)\left( z^l (\mathcal{C}_{q_{2n}})^{m-\left[\frac{q_{2n}}{2}\right]} (\mathcal{S}_{q_{2n}}(z))^r \oplus 0_{q_{2n-1}} \right) \label{scpecnabla1def1}\\
&&\nabla_1 \left( 0_{q_{2n}} \oplus z'^{l'} (\mathcal{S}_{q_{2n-1}}(z'))^{m'} (\mathcal{\bar{C}}_{q_{2n-1}})^{r'-\left[\frac{q_{2n-1}}{2}\right]} \right) = \nonumber\\
&&=2\pi\ii \left( l' q_{2n-1}+m' \right) \left( 0_{q_{2n}} \oplus z'^{l'} (\mathcal{S}_{q_{2n-1}}(z'))^{m'} (\mathcal{\bar{C}}_{q_{2n-1}})^{r'-\left[\frac{q_{2n-1}}{2}\right]} \right) \label{scpecnabla1def2}\ .
\end{eqnarray}
Because $\left( l' q_{2n-1}+m' \right)$ takes all possible values from $\mathbb{Z}$ for the allowed values of $l'$ and $m'$, we see that the spectrum of the deformed derivative is exactly the same as in the non-deformed case (\ref{specnabla1})
\begin{eqnarray}\label{specnabla2}
Spec (\nabla_1) = 2\pi\ii \mathbb{Z} \ .
\end{eqnarray}
The very important difference is the degeneracy. Though the spectrum is infinite in both deformed and non-deformed cases, in the deformed one it seems to be doubled for the lower part of the spectrum. Let $\lambda \in Spec (\nabla_1)$ and $|\lambda|< \left[\frac{q_{2n}}{2}\right]$, then both types of the eigen-vectors, (\ref{scpecnabla1def1}) and (\ref{scpecnabla1def2}) will contribute. While for $\lambda \in Spec (\nabla_1)$ and $|\lambda|> \left[\frac{q_{2n}}{2}\right]$, since $m$ has a finite range, only the vectors (\ref{scpecnabla1def2}) will correspond to the eigen-values from this part of the spectrum. Let us discuss the two cases in turn.

$|\lambda|< \left[\frac{q_{2n}}{2}\right]$. We can take two mutually orthogonal linear combinations of the eigen-vectors (\ref{scpecnabla1def1}) and (\ref{scpecnabla1def2}): one being the sum of two vectors with $\left( m-\left[\frac{q_{2n}}{2}\right] \right) = \left( l' q_{2n-1}+m' \right)$ (this is always uniquely satisfied for the specified above values of the indices) and the other their difference (with some relative coefficient). It is immediately clear that the sum corresponds to $\Gamma_n (\mathbf{u}^k \mathbf{v}^s)$ with $k = \left( m-\left[\frac{q_{2n}}{2}\right] \right)$ and arbitrary $s$ (which is related in some unique way to $l,r,r'$). These are exactly the eigen-vectors in the non-deformed case.\footnote{The appearance of the extra degeneracy compared to the commutative case is due to the ``difference'' eigen-vectors. It seems to be an artifact of the too naive choice of the deformed derivatives (note that in the absence of Leibnitz rule there is large freedom in defining these derivatives and some additional guiding principle is required). Though this will not effect our analysis of the spectral dimension, this point, i.e. how these vectors disappear/decouple in the continuous limit, should be clarified before the future applications of this noncommutative space could be discussed.}

$|\lambda|> \left[\frac{q_{2n}}{2}\right]$. This is the UV part of the spectrum and this is why the eigen-vectors (\ref{scpecnabla1def2}) have this strange form - this part would be pushed away in the $q_{2n} \rightarrow \infty$ limit.

Now the analysis of the spectrum of the deformed Laplacian (\ref{deformed_Laplace}) is almost trivial. Clearly the eigen-vectors (\ref{scpecnabla1def1}) and (\ref{scpecnabla1def2}) will continue to be the eigen-vectors of $\Delta_{(n)}$. The spectrum now will be given by
\begin{eqnarray}\label{specnablaLaplace}
Spec (\Delta_{(n)}) &=& \left\{ 4\pi^2 (k^2 + s^2)\ ,\ k\in \overline{-\left[\frac{q_{2n}}{2}\right],\left[\frac{q_{2n}}{2}\right]}\ ,\ s\in \mathbb{Z}\right\}\bigcup \nonumber \\
&&\bigcup\left\{ 4\pi^2 (k'^2 + s'^2)\ ,\ k'\in \overline{-\left[\frac{q_{2n-1}}{2}\right],\left[\frac{q_{2n-1}}{2}\right]}\ ,\ s'\in \mathbb{Z}\right\} \ .
\end{eqnarray}

The spectrum has a well-controlled behaviour: below the UV cut-off set by $q$ it has exactly the form of the spectrum for the usual torus  and above UV cut-off it goes as a spectrum for two copies of $\mathbb{S}^1$ (see the next section). As in the examples of Section \ref{spectvsscal} we see that there is a cutoff, which controls the dimensional behaviour of the space.

\setcounter{equation}{0}
\section{Spectral dimension}\label{se:spectral}

In this section we will study the spectral dimension of fuzzy geometry defined above. We proceed along the lines of Sect.~\ref{spectvsscal}., i.e.\ we use the generalization of Weyl theorem applied to the spectrum (\ref{specnablaLaplace}) of the deformed Laplacian to define the spectral dimension of our fuzzy torus. We want to calculate the spectral dimension of our fuzzy geometry in two extreme limits, infrared and ultraviolet. Even before performing the actual calculation we can make some comments on what one should expect to see in these limits. As it was discussed in the beginning of the paper, the physical spectral dimension is the dimension as seen in the experiment that can probe the geometry up to some cut-off scale. This means that the IR limit should look as the commutative geometry, i.e.\ we expect that the spectral dimension in this case should be 2. In the ultraviolet (UV) limit we do not have, in general, enough intuition (which is based on a commutative geometry). In this case the actual calculation should provide us with some hints on where the fundamental, i.e.\ UV, degrees of freedom really live. We will see that this is the case.

Before we proceed to a more detailed analysis it is instructive to compare the spectrum~(\ref{specnablaLaplace}) with the one for a torus with one fuzzy dimension (\ref{spec_nc}) (for the case $R\sim r$). It is obvious that modulo some finite degeneracy (see the footnote above) these spectra are essentially the same. But it is also clear that the \textit{finite} degeneracy cannot change the spectral growth (it will only change a ``volume'' prefactor in the generalized Weyl's formula). Therefore, we expect for the spectral dimension the same behaviour as schematically depicted on Fig.2. Let us see this in more detail.

\textbf{IR Regime}. As we discussed above, by IR we mean that the cut-off scale $\omega$ is below the characteristic quantum geometric scale. In the case of a toy model this scale was controlled by the number of the states along $R$-direction. In the present case, this means that $\omega < q_{2n-1}^2$, it does not even have to be much smaller. It follows that only the winding modes (from two circles) with $l,l' = -1,0$ contribute. Then we immediately have for the counting function (compare with (\ref{counting_torus}))
\begin{eqnarray}\label{counting_IR}
N_\Delta (\omega) \sim \mbox{degeneracy} \times\int\limits_{m^2 + s^2 \leq \frac{\omega}{4\pi^2}} d m\, d s = \mbox{const} \times \omega \ .
\end{eqnarray}
Applying our definition of the dimension (\ref{scaling_dim}) we immediately get $d_{\mbox{\scriptsize IR}} = 2$.

As we discussed, this result is not unexpected. On the technical side, this is the consequence of the fact that the \textit{effective} radii of two $\mathbb{S}^1$ in $\mathrm{Mat}_q (\mathcal{C}^\infty (\mathbb{S}^1)) \oplus \mathrm{Mat}_q (\mathcal{C}^\infty (\mathbb{S}^1))$ are very small. Although we started with all the radii of the order of 1 (we are working with the dimensionless radii), the contribution of ($l,l'$)-mode to the spectrum is of the order of $q^2\gg 1$ (where $q$ is either $q_{2n}$ or $q_{2n-1}$), see (\ref{scpecnabla1def2}). This effectively reduces the radii of the ``internal'' circles by the factor of $q$, making them ``unobservable'' in IR.

\textbf{UV Regime}. This is the case opposite to the previous one, i.e.\ many of the $\mathbb{S}^1$ winding modes are excited, $l,l' \gg 1$. This means that the hypothetical experiment can probe the physics up to the cut-off $\omega \gg q_{2n}^2 , q_{2n-1}^2$. In this case we have for the spectrum (we use the representation in terms of $l',m',r'$, see the discussion after (\ref{specnabla2}))
\begin{eqnarray}\label{spectrum_UV}
4\pi^2\left((r'-\left[\frac{q_{2n-1}}{2}\right])^2 + (q_{2n-1} l'+m')^2\right) = 4\pi^2 q_{2n-1}^2 l'^2 \left(1 + \mathcal{O}\left(\frac{1}{l'}\right)\right) \ ,
\end{eqnarray}
where we used that $r' , m' \in [0, q_{2n-1})$, see (\ref{truncation1}).
Then we can write for the counting function in this limit\footnote{Of course, the same should be done for the other set of vectors (\ref{scpecnabla1def1}), but for the regime when $\omega\gg q_{2n},q_{2n-1}$ it will essentially produce the same result (\ref{counting_UV}) with $q_{2n}$ as a factor, reflecting the existence of two circles.}
\begin{eqnarray}\label{counting_UV}
N_\Delta (\omega) \rightarrow \mbox{degeneracy} \times \iint\limits^{\ \ q_{2n-1}} d m\, d r \!\!\!\int\limits_{-\frac{\sqrt{\omega}}{2\pi q_{2n-1}}}^{\frac{\sqrt{\omega}}{2\pi q_{2n-1}}}\!\!\! d k = \mbox{const}\times q_{2n-1}\sqrt{\omega} \ .
\end{eqnarray}
Again, applying (\ref{scaling_dim}) we get the physical dimension in ultraviolet $d_{UV} = 1$. The exact form of the scaling dimension $d(\omega)$ will be similar to (\ref{scaling_dim2b}). We intensionally left the factor of $q$ in (\ref{counting_UV}). Recalling the original Weyl theorem (\ref{Weyl}) we see that the effective size of the UV-dimension is proportional to $q$ instead of being order one, or even being of order of $1/q$ (recall that the effective radii of $\mathbb{S}^1$ are reduced be the factor of $q$). This ``lengthening'' is due to the matrix degrees of freedom, namely, the fact that there are a number of order of $q^2$ of them. This is a very important result: in ultraviolet the new dimension (coming from two $\mathbb{S}^1$, i.e.\ \textit{not} related to the IR dimensions) appears to be fundamental and the IR dimensions of the commutative torus $\mathbb{T}^2$ disappear completely, the only trace of their presence being the lengthening of the UV-dimension (which happens at the expense of the complete loss of the IR dimensions). The deception has been unmasked!

\setcounter{equation}{0}
\section{Discussion and conclusions}

We have shown that it is possible to have a space for which the number of dimensions can be different depending on whether it is probed at high energies (short distances) or low energies (large distances). The UV and IR geometries are quite different: while in IR regime the geometry appears to be a 2d torus, in UV it results being two disconnected lines (or circles whose length goes to infinity as $q_{2n}$ or $q_{2n-1}$). Although our model is two dimensional, higher dimensional versions of the Elliot-Evans construction are possible \cite{QLin, NCPhilllips} and a construction similar to the one performed in this paper can be done in more generality, with the high energy space being composed by an higher number of circles (or, possibly, tori).

Of course, the model presented here is not realistic, but it shows that by allowing space-time to have a nontrivial quantum/noncommutative structure the interplay between long and short distances (high and low energies) may produce a rich structure for which the number of dimensions is changed, yet the isometries of the original space are preserved, in particular there is still no preferred direction. We note that the noncommutativity of the space can be arbitrarily small, in fact it is possible to have the construction in such a way the $\lim_n\theta_n\to 0$. In this way locality would be preserved.

Our model is too simple, yet it still can be used to study some novel phenomena due to quantum structure of space-time. In this respect, there are several very important issues to be addressed within our model. First of all, one has to better understand the nature of the extra degeneracy in the spectrum, discussed in the sections \ref{se:derivatives} and \ref{se:spectral}. While, as we stressed, it does not effect our analysis or conclusions, to have a better control on the geometrical aspects of the quantum space-time this point needs clarification. Most probable, this would require a more careful study of all possible (approximate) derivations of the algebra of a fuzzy torus, that are natural in the sense that they respect the direct sum structure. This point is tightly connected with the next step: construction of a Dirac operator (and not just Laplacian).

The role of the Dirac operator for the physical models based on the noncommutative geometry is two-folded: firstly, it controls the geometry of the underlying space-time and we have partially addressed this in our work (namely the dimensional aspect of our model); secondly, it \textit{defines} the dynamics of the matter sector via the so-called spectral action \cite{Chamseddine:1996zu,Chamseddine:2008zj}.
This second role of the Dirac operator should be very interesting to study for our model. In principle, it should allow the explicit analysis of the microscopic, UV, dynamics of the matter fields. Based on the general arguments of the present work, it is clear that this dynamics will be completely different from the effective, IR, one, revealing the ``true'' degrees of freedom. The effectiveness of the spectral action approach for the class of models with a ``flowing'' spectral dimension, the so-called Horava-Lifshitz models, was demonstrated in \cite{Lopes:2015bra,Pinzul:2016dwy}. It was shown that the spectral action severely restricts the parameter space of the matter sector by introducing the strong dependence between parameters in gravity and matter sides. As we mentioned, in our model the mechanism of the dimensional flow is quite different from the Horava-Lifshitz one. So, it would be very interesting to compare these two approaches on the level of matter dynamics. Another related issue is the following one: The spectral action is a residue coming from a heat kernel expansion~\cite{Vassilevich:2003xt, Connes:2006qj, Eckstein:2015rya}, but can be obtained also from cancellation of anomalies~\cite{Andrianov:1983fg, Andrianov:1983qj, Andrianov:2010nr, Kurkov:2012dn} or a $\zeta$-function regularization~\cite{Kurkov:2014twa}. The presence of spaces, such as the one described here, with a built-in cutoff, alter profoundly the field theory, and in particular the UV dynamics of bosons~\cite{Kurkov:2013kfa, Alkofer:2014raa}. It would be interesting to investigate the fate of field theory on these spaces using spectral tools and/or asymptotic safety~\cite{Niedermaier:2006wt}. We hope to address this and other questions elsewhere.

\appendix
\setcounter{equation}{0}
\section{The Elliot-Evans construction}

In this appendix we describe the matrix approximation to the noncommutative torus used in the paper. In particular we wish to describe in detail what kind of a fuzzy torus we used in our analysis. As mentioned above, to construct a noncommutative geometry one needs several ingredients. To define the topological part of a noncommutative space, we need an algebra (of ``continuous'' functions), while Dirac operator (essentially the notion of derivatives) is responsible for the geometry. Here we construct the answer to the first part, i.e. an algebra, while the ``geometrical'' part, i.e.\ derivatives, is addressed in section~\ref{se:derivatives}. It is done via the Elliott-Evans (EE) construction~\cite{Elliott:1993}. Because this construction is in the heart of our work, and yet is relatively unknown to a broader community, we review it in some details. We shall see that the noncommutative torus can be rigorously approximated by an algebra of matrices of functions on the one dimensional space - a topological sum of two circles. This construction will be expedient for the truncation which we will perform in the section \ref{se:truncation} to further obtain the scaling dimension.

The EE result is the constructive answer to the following question: Can we approximate the algebra of a noncommutative torus, $\mathcal{A}_\theta$ defined in (\ref{algebraAtheta}), by some finite dimensional, i.e.\ matrix, algebras? In section \ref{se:approximation} we saw how using the simple \textit{K}-theoretical arguments one can show that the most natural guess - an approximation by the inductive limit of AF-algebras - does not work.  We will see how the EE construction overcomes this \textit{K}-theoretic obstruction.

The basic idea of EE construction is as follows: find the approximation of the generators of the noncommutative algebra $\mathcal{A}_\theta$, $\mathbf{vu}=\omega \mathbf{uv}$,  $\omega=\exp (2\pi\ii\theta)$ by some tower of projectors in $\mathcal{A}_\theta$. The construction is based on some generalization of the Rieffel projection \cite{Rieffel:1981}. Let us briefly recall what it is.

Let $\mathbf{f}$ and $\mathbf{g}$ be some elements of $C^* (\mathbf{u})\simeq \mathcal{C}(\mathbb{T})$ (continuous functions on a circle) to be defined later and $q'$ be some positive integer. Then define an element of $\mathcal{A}_\theta$\footnote{The original Rieffel's construction was for $q'=1$.}
\begin{eqnarray}\label{projector11}
P_{11} := \mathbf{v}^{-q'} \mathbf{g} + \mathbf{f} + \mathbf{g} \mathbf{v}^{q'} \ .
\end{eqnarray}
To (almost) fix the elements $\mathbf{f}$ and $\mathbf{g}$ we require that
\begin{itemize}
\item[1)] $P_{11}\in \mathcal{A}_\theta$ is a projector;
\item[2)] $\mathrm{Tr} P_{11} = p' - q'\theta =: \beta $, i.e.\ $P_{11}$ represents the $(p' ,-q')$-class in $K_0$. Here $p'$ is some integer, such that $\beta \in [0,1]$\ .
\end{itemize}
These conditions determine the elements $\mathbf{f}$ and $\mathbf{g}$. Namely,
\begin{itemize}
\item[1)] $f$ is a continuous function with $\mathrm{supp} f \in[0, 1/q]$, here $q$ is an integer defining the number of the projectors in the tower (see below);
\item[2)] $\mathbf{g}\mathbf{v}^{-q'}\mathbf{g}\mathbf{v}^{q'}=0$, $(\mathbf{f} + \mathbf{v}^{q'}\mathbf{f}\mathbf{v}^{-q'})\mathbf{g}=\mathbf{g}$, $\mathbf{g} + \mathbf{v}^{-q'}\mathbf{g}\mathbf{v}^{q'}= (\mathbf{f}-\mathbf{f}^2)^{1/2}$\ ;
\item[3)] $\int\limits_0^1 f(x)\, dx = \beta$\ .
\end{itemize}
Here $f(x)\in \mathcal{C}(\mathbb{T})$ is related to $\mathbf{f}\in C^* (\mathbf{u})$ by the continuous functional calculus:
\begin{eqnarray}\label{spectral}
\mathbf{f} = \int\limits_0^1 f(
x)d E_x\ , \ \mathrm{where}\ \ \mathbf{u} = \int\limits_0^1 \mathrm{e}^{2\pi\ii x} d E_x \ ,
\end{eqnarray}
i.e.\ $E_x\ , \ x\in [0,1)\equiv \frac{1}{2\pi\ii}\ln\mathrm{Spect}(\mathbf{u})$, is the family of the spectral projections corresponding to $\mathbf{u}$ and we denoted $f(\mathrm{e}^{2\pi\ii x})$ just by $f(x)$. The actual shape of $f(x)$ is further restricted by the condition that the accuracy of the approximation we are looking for would be the best possible. This requires that
\begin{itemize}
\item[4)] the slopes of the non-constant parts of $f(x)$ are minimal possible.
\end{itemize}
We will not need the further details about $f(x)$, see \cite{Elliott:1993} and \cite{Landi:2004sc} for an explicit example (including some plots of the function).

To construct the whole tower we employ the canonical action of the torus $\mathbb{T}^2$ on the noncommutative torus $\mathcal{A}_\theta$. Given a point on $\mathbb T^2$ consider the action on a monomial from $\mathcal A_\theta$ as $\alpha_{z_1 z_2}(\mathbf{u}^n\mathbf{v}^m ) = z_1^n z_2^m \mathbf{u}^n \mathbf{v}^m$  $\forall (z_1 , z_2)\in \mathbb{T}^2$. Fix an integer $p$ relatively prime with $q$ and define $P_{ii},\ i=\overline{1,q}$ by
\begin{eqnarray}\label{projectorii}
P_{ii} := (\alpha_{\mathrm{e}^{2\pi\ii p/q},1})^{i-1} (P_{11}) = :\alpha^{i-1} (P_{11})\ .
\end{eqnarray}
It is pretty straightforward to see that the $P_{ii}$ are the projectors and the choice of the support of $\mathbf{f}$ in (\ref{projector11}) guarantees that they are orthogonal, $P_{ii} P_{jj} = \delta_{ij} P_{ii}$ (no sum over $i,j$). Let us denote $\tilde{\mathbf{r}} = \alpha (\mathbf{r})$ for any element  $\mathbf r\in C^*(\mathbf{u})$. Because the effect of applying $\alpha$ is the translation of the spectral support of $\mathbf{r}$ by $p/q$, we see that if $\mathrm{supp}(\mathbf{r})\subset [0,1/q]$ then $\mathrm{supp}(\tilde{\mathbf{r}})\subset [p/q,(p+1)/q]$, so $\mathrm{supp}(\tilde{\mathbf{r}}) \cap \mathrm{supp}(\mathbf{r}) = \emptyset$, i.e.\ $\mathbf{t}\,\tilde{\mathbf{r}}=0$, where $\mathbf{t}$ is an arbitrary element with the spectral support in $[0,1/q]$. Using this and that $\mathbf{v}^{-q'}\mathbf{g}\mathbf{v}^{q'}=(\mathbf{f}-\mathbf{f}^2)^{1/2} - \mathbf{g}$ (i.e.\ it also has a spectral support in $[0,1/q]$), it is a trivial exercise to show that $P_{11} \alpha (P_{11})=0$ or $P_{ii} P_{jj} = \delta_{ij} P_{ii}$ in general.

The importance of this tower of projectors is due to the following estimates \cite{Elliott:1993}
\begin{eqnarray}\label{approx1}
\| \mathbf{u} P_{11} \mathbf{u}^{-1} - P_{11}\| < C(q,q')\frac{1}{q}\ , \\
\| \mathbf{v} P_{11} \mathbf{v}^{-1} - \alpha (P_{11})\| < C(q,q')\frac{1}{q}\ ,
\end{eqnarray}
where $C(q,q')$ is some bounded function whose explicit form is irrelevant for us. While $\mathbf{u}$ almost commutes with $P_{11}$ (when $q,q' \rightarrow\infty$), the adjoint action of $\mathbf{v}$ approximately reproduces the representation of $\alpha$. Due to the trivially verified property $\alpha^q = 1$, we see that the height of the tower of the projectors (\ref{projectorii}) is exactly given by $q$ and that $P:=\sum\limits_{i=1}^q P_{ii}$ approximately commutes with both $\mathbf{u}$ and $\mathbf{v}$. This is exactly the result that is crucial for the EE approximation.

Instead of the estimates (\ref{approx1}) we will need the slightly modified ones:
\begin{eqnarray}\label{approx2}
\| \mathbf{u} P_{11} - P_{11}\| < C(q,q')\frac{1}{q}\ , \\
\| \mathbf{v} P_{11} - \alpha (P_{11}) \mathbf{v} P_{11}\| < C(q,q')\frac{1}{q}\ ,
\end{eqnarray}
with possibly different function $C(q,q')$. E.g.\ let us demonstrate (\ref{approx2}). Both, $f(x)$ and $g(x)$, have the support in $[0,1/q]$. Then for any element $\mathbf{f}$ in $C^* (\mathbf{u})$ corresponding to such a function and an arbitrary element $\mathbf{r}$ in $C^* (\mathbf{v})$ we have
\begin{eqnarray}
\|\mathbf{u}\mathbf{f}\mathbf{r} - \mathbf{f}\mathbf{r}\| \leq \|\int\limits_0^1 (\mathrm{e}^{2\pi\ii x} - 1) f(x)d E_x \| \|\mathbf{r} \| = \sup\limits_{x\in[0,1/q]} |(\mathrm{e}^{2\pi\ii x} - 1) f(x)| \|\mathbf{r} \| \leq  \frac{2\pi}{q}\|\mathbf{f}\| \|\mathbf{r} \|\ .
\end{eqnarray}
Using this result and the definition of $P_{11}$ (\ref{projector11}) we get
\begin{eqnarray}
\| \mathbf{u} P_{11} - P_{11} \| = \| \mathbf{e}^{-2\ii \pi q' \theta} \mathbf{v}^{-q'}\mathbf{g} (\mathbf{e}^{-2\ii \pi q' \theta} \mathbf{u}-1) + (\mathbf{u}-1)(\mathbf{f}+\mathbf{g} \mathbf{v}^{q'})\| \leq \frac{7\pi}{q} \ ,
\end{eqnarray}
where we used $\beta<1/2$ and $\|\mathbf{g}\| = 1/2$, which is a trivial consequence of the relation between $\mathbf{g}$  and $\mathbf{f}$ (and in any case it is not that important for establishing (\ref{approx2}) as long as $\|\mathbf{g}\|$ is finite). The other estimates can be obtained in an analogous manner.

A second tower is necessary for the approximation since so far we treated $\mathbf u$ and $\mathbf v$ not symmetrically. Let us look at the trace of $P:=\sum\limits_{i=1}^q P_{ii}$:
\begin{eqnarray}
\mathrm{Tr} P = q\beta , \end{eqnarray}
where $q$ comes from\ the\ height\ of\ the\ tower\ and\ $\beta$ is\ the\ trace\ of\ each\ individual\ $P_{ii}$.
Then we see that if we really want to have an approximation to $\mathcal{A}_\theta$, the \textit{K}-theoretic argument requires the second tower $P'$ of the trace $\mathrm{Tr}P' = 1-q\beta$. Requiring the height of this tower to be $q'$ and the trace of each projector $\beta'$, we get the condition
\begin{eqnarray}\label{qqpp}
q\beta + q'\beta' = 1 ,\ \mathrm{which\ is\ solved\ by\ }\beta' = q\theta - p\ ,
\end{eqnarray}
i.e.\ the 2 by 2 matrix $\left(
                          \begin{array}{cc}
                            p' & p \\
                            q' & q \\
                          \end{array}
                        \right)
$ is an element of $SL(2,\mathbb{Z})$ group. Because the defining relations of $\mathcal{A}_\theta$ are invariant under the automorphism: $\mathbf{u}\mapsto \mathbf{v}$ and $\mathbf{v}\mapsto \mathbf{-u}$, the construction of the second tower seems straightforward:
\begin{eqnarray}\label{projectorprime11}
P'_{11} := \mathbf{u}^{q} \mathbf{g'} + \mathbf{f'} + \mathbf{g'} \mathbf{u}^{-q} \ .\\
P'_{ii} := \alpha'^{i-1}(P'_{11})\ ,i=\overline{1,q'} \ ,
\end{eqnarray}
where $\mathbf{g'}$ and $\mathbf{f'}$ are now elements of $C^*(\mathbf{v})$ and $\alpha'$ is the action on $\mathcal{A}_\theta$ by the element of $\mathbb{T}^2$ $\alpha_{1,\mathrm{e}^{-2\pi\ii p'/q'}}$. It is obvious that all the relations satisfied by $P_{ii}$ will be true for $P'_{ii}$ after exchanging $q$ and $q'$. So we can sum up our estimates for the operators generating these two towers\footnote{Again, possibly with a different function $C(q,q')$.}:
\begin{eqnarray}
\| \mathbf{u} P_{11} - P_{11}\| < C(q,q')\mathrm{max}\left( \frac{1}{q}, \frac{1}{q'}\right)\ , \label{approxtotal1}\\
\| \mathbf{v} P_{11} - \alpha (P_{11})\mathbf{v} P_{11}\| < C(q,q')\mathrm{max}\left( \frac{1}{q}, \frac{1}{q'}\right)\ ,\label{approxtotal2}\\
\| \mathbf{v} P'_{11} - P'_{11}\| < C(q,q')\mathrm{max}\left( \frac{1}{q}, \frac{1}{q'}\right)\ , \label{approxtotal3}\\
\| \mathbf{u} P_{11} - \alpha' (P'_{11})\mathbf{u} P'_{11}\| < C(q,q')\mathrm{max}\left( \frac{1}{q}, \frac{1}{q'}\right) \label{approxtotal4}\ .
\end{eqnarray}
There is a problem though. While the projectors within each tower are mutually orthogonal, $P$ and $P'$ are not! Fortunately, \textit{K}-theoretic argument again shows that there should exist a unitary operator $\mathbf{W}$ that takes $P'$ to the orthogonal complement of $P$.\footnote{Because two projectors, $P'$ and $1-P$, have the same trace, $1-q\beta$, they should be unitary equivalent.} What is slightly less trivial is that this unitary $\mathbf{W}$ can be chosen in such a way that it approximately commutes with $\mathbf{u}$ and $\mathbf{v}$ so the estimates (\ref{approxtotal1}-\ref{approxtotal4}) will not be spoiled. We will not show this, for the details see \cite{Elliott:1993}.

Now we are ready to construct the approximation. We consider $p,q,p',q'$ large with $p/q\sim p'/q' \sim \theta$. To begin with, note that $\alpha^i (\mathbf{u} P_{11} - P_{11})\alpha^j (\mathbf{u} P_{11} - P_{11})=0$ for $i\neq j$. This is trivially shown using the same considerations we have made to demonstrate that $P_{11}$ and $\alpha (P_{11})$ are orthogonal projectors. Then noting that $\alpha (\mathbf{u}) = \xi\mathbf{u}$,  where $\xi = \exp (2\pi\ii p/q)$ we have (recall that $P$ is the sum of all the projectors in the first tower):
\begin{eqnarray}\label{approxu1}
\left \| \mathbf{u} P -\sum\limits_{k=0}^{q-1} \xi^{-k} P_{kk}\right\| = \left \| \sum\limits_{k=0}^{q-1}\xi^{-k}\alpha^{k}(\mathbf{u} P_{11} - P_{11})\right\| &=& \sup\limits_k \left \| \xi^{-k}\alpha^{k}(\mathbf{u} P_{11} - P_{11})\right\| \nonumber\\ &\le& C(q,q')\mathrm{max}\left( \frac{1}{q}, \frac{1}{q'}\right) \ ,
\end{eqnarray}
where the second equality is possible exactly due to the fact that all the terms in the sum have non-overlapping supports/ranges and at the end we used (\ref{approxtotal1}).

Repeating the same with the second tower and $\mathbf{v}$ instead of $\mathbf{u}$, we get ($\alpha ' (\mathbf{v}) = \xi'^{-1}\mathbf{v}$, where $\xi' = \exp (2\pi\ii p'/q')$) \footnote{In the paper \cite{Elliott:1993} there are some sign and notational errors for this estimate, which do not affect the conclusions of that paper but are important for us.}
\begin{eqnarray}\label{approxv1}
\left \| \mathbf{v} P' -\sum\limits_{k=0}^{q'-1} \xi'^{k} P'_{kk}\right\| = \left \| \sum\limits_{k=0}^{q'-1}\xi'^{k}\alpha'^{k}(\mathbf{v} P'_{11} - P'_{11})\right\| &=& \sup\limits_k \left \| \xi'^{k}\alpha'^{k}(\mathbf{v} P'_{11} - P'_{11})\right\| \nonumber\\ &\le& C(q,q')\mathrm{max}\left( \frac{1}{q}, \frac{1}{q'}\right) \ .
\end{eqnarray}
(\ref{approxu1}) and (\ref{approxv1}) are our first two out of four the most important estimates. They show that while $\mathbf{u}$ is the best approximated on the range of $P$ and this approximation is given by
\begin{eqnarray}\label{Cq}
\mathcal{C}_q := \sum\limits_{k=0}^{q-1} \xi^{-k} P_{k+1,k+1} \ ,
\end{eqnarray}
the subspace of the best approximation for $\mathbf{v}$ is the range of $P'$\footnote{This \textit{is} the main reason, why one needs two towers!} with the approximation
\begin{eqnarray}\label{Cq1}
\mathcal{\bar{C}}_{q'} := \sum\limits_{k=0}^{q'-1} \xi'^{k} P'_{k+1,k+1} \ .
\end{eqnarray}
Note that though the approximations are finite, i.e.\ given in terms of the finite number of the projectors, they still belong to the full algebra of the noncommutative torus, $\mathcal{A}_\theta$.

Of course, now we would like to see what are the best approximations for $\mathbf{v}$ ($\mathbf{u}$) on the domain of $P$ ($P'$). Unfortunately, the answer to this question is slightly more complicated than our previous consideration. We will deal with the case of $\mathbf{v}$ in details, while the other case is completely identical (with the obvious interchange of primed and unprimed quantities).

Clearly, now we want to work with the estimate (\ref{approxtotal2}). The first problem is due to the range of the element $\mathbf{v}P_{11} - \alpha (P_{11})\mathbf{v} P_{11}$. While the range of the second term is contained in the range of $P_{22}\equiv \alpha (P_{11})$ (i.e.\ corresponds to the spectral support inside $[p/q,(p+1)/q)$, see (\ref{spectral})) the range of the first element is supported inside $[\theta , 1/q +\theta )$. It can be shown that this has a non-trivial overlap with $[(p+1)/q ,(p+2)/q)$. In fact
\begin{eqnarray}
\frac{p}{q}< \theta < \frac{p'}{q'}\Rightarrow \frac{p+1}{q}< \theta + \frac{1}{q} < \frac{p'q + q'}{qq'}=\frac{1 + pq' + q'}{qq'}= \frac{p+1}{q} + \frac{1}{qq'} < \frac{p+2}{q} \ ,
\end{eqnarray}
where we used that $qp' - q'p = 1$. This means that now we do not have in general $\alpha^i (\mathbf{v}P_{11} - \alpha (P_{11})\mathbf{v} P_{11}) \alpha^j ( \mathbf{v}P_{11} - \alpha (P_{11})\mathbf{v} P_{11}) = 0$ for $i\neq j$ as before (which was crucial for the last equality in (\ref{approxu1})). But the same consideration shows that we have two families: the orbits of $\mathbf{v}P_{11} - \alpha (P_{11})\mathbf{v} P_{11}$ under the \textit{even} number of actions by $\alpha$ and the orbits of $\mathbf{v}P_{11} - \alpha (P_{11})\mathbf{v} P_{11}$ under the \textit{odd} number of actions by $\alpha$. Within each family the elements have mutually orthogonal supports/ranges. Then we can write the estimate analogous to (\ref{approxu1}) (recall that $\alpha (\mathbf{v}) = \mathbf{v}$)
\begin{eqnarray}\label{approxv2}
&&\left \| \mathbf{v} P -\sum\limits_{k=0}^{q-1} \alpha^k ( \alpha (P_{11})\mathbf{v} P_{11} )\right\| = \left \| \sum\limits_{k=0}^{q-1}\alpha^{k}(\mathbf{v} P_{11} - \alpha (P_{11})\mathbf{v} P_{11} )\right\| = \nonumber\\
 &=& \left \| \sum\limits_{k\ even}\alpha^{k}(\mathbf{v} P_{11} - \alpha (P_{11})\mathbf{v} P_{11} ) + \sum\limits_{k\ odd}\alpha^{k}(\mathbf{v} P_{11} - \alpha (P_{11})\mathbf{v} P_{11} )\right\| \le\nonumber\\
&\le& \left \| \sum\limits_{k\ even}\alpha^{k}(\mathbf{v} P_{11} - \alpha (P_{11})\mathbf{v} P_{11} ) \right\|+ \left\| \sum\limits_{k\ odd}\alpha^{k}(\mathbf{v} P_{11} - \alpha (P_{11})\mathbf{v} P_{11} )\right\| \le \nonumber \\
&\le& 2 C(q,q')\mathrm{max}\left( \frac{1}{q}, \frac{1}{q'}\right) \ .
\end{eqnarray}
So we see that the same estimate is still valid (the factor of 2 is irrelevant because it always can be re-absorbed into the definition of the bounded function $C$.) Though (\ref{approxv2}) already looks like an approximation for $\mathbf{v}$ on the range of $P$, it cannot be taken as satisfactory because it is explicitly defined through $\mathbf{v}$ itself. We would like to construct an approximation in terms of some fixed elements of the algebra $\mathcal{A}_\theta$ defined by the system of the projectors (as in (\ref{approxv1}), which is given in terms of the projectors only). This is the second complication in the construction.

To proceed, let us note that the building block of the approximation (\ref{approxv2}), $\alpha (P_{11})\mathbf{v} P_{11} \equiv P_{22}\mathbf{v} P_{11}$, maps from the support of $P_{11}$ into the range of $P_{22}$, but this map is not isometric. Thinking of $P_{22}\mathbf{v} P_{11}$ as a bounded operator (in the GNS construction) we can always write a polar decomposition
\begin{eqnarray}\label{P21}
P_{22}\mathbf{v} P_{11} = P_{21} | P_{22}\mathbf{v} P_{11} | \ ,
\end{eqnarray}
where $P_{21}$ is a partial isometry (or a unitary operator from the support of $P_{11}$ to the range of $P_{22}$). The key observation (that can be shown using the same methods as above) is that this partial isometry is almost ``the same'' as $P_{22}\mathbf{v} P_{11}$ itself
\begin{eqnarray}\label{approxP21}
\| P_{22}\mathbf{v} P_{11} -  P_{21} \| \le C(q,q')\mathrm{max}\left( \frac{1}{q}, \frac{1}{q'}\right)\ .
\end{eqnarray}
Now, in complete analogy with (\ref{projectorii}), we construct a tower of partial isometries
\begin{eqnarray}\label{projectorisometry}
P_{2+i,1+i} := \alpha^{i} (P_{21})\ ,\ i=\overline{0,(q-2)}\ .
\end{eqnarray}
It is not hard to see that by the construction $P_{ii}$ and $P_{2+i,1+i}$ have orthogonal ranges and supports, i.e.\
\begin{eqnarray}\label{matrixunits}
P_{ii}P_{jj}=\delta_{ij}P_{ii}\ , \ P_{2+i,1+i}P_{jj} = \delta_{1+i,j}P_{j+1,j}\ \mathrm{and}\ P_{jj}P_{2+i,1+i} = \delta_{2+i,j}P_{j,j-1}\ .
\end{eqnarray}
Also $P_{ii}$ satisfy
\begin{eqnarray}\label{trace}
\sum\limits_{i=1}^{q}P_{ii} = P =: \mathds{1}_q \ .
\end{eqnarray}
One can recognize in (\ref{matrixunits}) the part of the defining relations of the $q\times q$ matrix units (namely, the relations between $2q-1$ of them), see e.g. \cite{Lam:1998}. The full set of the defining relations is given by
\begin{eqnarray}\label{matrixunits1}
\forall i,j = \overline{1,q}\ \ P_{ij}P_{ls}=\delta_{jl}P_{is}\ , \ \sum\limits_{i=1}^{q}P_{ii} = P =: \mathds{1}_q \ .
\end{eqnarray}
We generate the remaining matrix units from the set of the projectors $P_{ii}$ and partial isometries $P_{2+i,1+i}$
\begin{eqnarray}\label{matrixunits2}
&&P_{ij} \ \mathrm{for}\ i>j \ \mathrm{is\ defined\ by}\ P_{ij}:=P_{i,i-1}P_{i-1,i-2}\cdots P_{j+1,j}\ , \\
&&\mathrm{while\ for}\ i<j\ \mathrm{we\ define} \ P_{ij}:= P_{ji}^\dagger\ .
\end{eqnarray}
Using the fact that $P_{i+1,i}$ are partial isometries (or isometries from the range of $| P_{22}\mathbf{v} P_{11} |$ to the range of $P_{22}\mathbf{v} P_{11}$) it is easy to see that the defining relations for matrix units (\ref{matrixunits1}) are satisfied.

We would like to use the estimate (\ref{approxv2}) to approximate $\mathbf{v}$ on the range of $P$ (and in complete analogy $\mathbf{u}$ on the range of $P'$) as it was done for $\mathbf{u}P$ and $\mathbf{v}P'$ using the estimates (\ref{approxu1}) and (\ref{approxv1}). Unfortunately this still cannot be done in terms of the matrix units only (i.e.\ in terms of the projectors and partial isometries). The problem is that the sum in the estimate (\ref{approxv2}) goes up to $q-1$, which produces the term whose approximation, as in (\ref{approxP21}), $\alpha^{q-1}(P_{21})$ is \textit{not} equal to any combination of the matrix units. But being the partial isometry with the same domain and range as $P_{1q}$ (see the comment after (\ref{matrixunits2})) it can differ from $P_{1q}$ only by a unitary on the range of $P_{11}$:
\begin{eqnarray}\label{unitaryz}
\alpha^{q-1}(P_{21}) = z P_{1q}\ ,
\end{eqnarray}
where $z$ is a unitary element in $P_{11}\mathcal{A}_\theta P_{11}$. We will see below that exactly this unitary element permits to circumvent the \textit{K}-theoretical obstruction for the finite dimensional approximations of $\mathcal{A}_\theta$.

Now we are finally in the position to finish our construction of the approximation of $\mathbf{v}$ on the range of $P$. Combining the estimates (\ref{approxv2}) and (\ref{approxP21})\footnote{And also using the orthogonality of the ranges and supports of $\alpha^i (P_{22}\mathbf{v}P_{11})$ and $\alpha^j (P_{21})$ for different $i$ and $j$ as it was done in, e.g. (\ref{approxu1}).}, we obtain
\begin{eqnarray}\label{approxv22}
&&\left \| \mathbf{v} P -\sum\limits_{k=0}^{q-2} P_{2+k,1+k} - z P_{1q} \right\| =\\
&& =\left \| \mathbf{v} P -\sum\limits_{k=0}^{q-1} \alpha^k ( \alpha (P_{11})\mathbf{v} P_{11} ) + \sum\limits_{k=0}^{q-1} \alpha^k ( \alpha (P_{11})\mathbf{v} P_{11} ) - \sum\limits_{k=0}^{q-1} \alpha^k (P_{21}) \right\| \le \nonumber\\
&& \le 2 C(q,q')\mathrm{max}\left( \frac{1}{q}, \frac{1}{q'}\right) + C(q,q')\mathrm{max}\left( \frac{1}{q}, \frac{1}{q'}\right) = 3 C(q,q')\mathrm{max}\left( \frac{1}{q}, \frac{1}{q'}\right) \ .
\end{eqnarray}

Trivially repeating the same consideration with the obvious changes, $\mathbf{v}\rightarrow\mathbf{u}$, $P_{ii}'$, $i=\overline{1,q'}$ and $\alpha \rightarrow \alpha'$, we obtain the estimate for $\mathbf{u}P'$
\begin{eqnarray}\label{approxu22}
&&\left \| \mathbf{u} P' -\sum\limits_{k=0}^{q'-2} P_{2+k,1+k}' - z' P_{1q'}' \right\| \le 3 C(q,q')\mathrm{max}\left( \frac{1}{q}, \frac{1}{q'}\right) \ ,
\end{eqnarray}
where $z'$ is now a unitary element in $P_{11}'\mathcal{A}_\theta P_{11}'$.

The estimates (\ref{approxv22}) and (\ref{approxu22}) show that the best approximation for $\mathbf{v}$ on the range of $P$ is given by
\begin{eqnarray}\label{Sq}
\mathcal{S}_q (z) := \sum\limits_{k=0}^{q-2} P_{2+k,1+k} + z P_{1q} \ ,
\end{eqnarray}
and the best approximation for $\mathbf{u}$ on the range of $P'$ is
\begin{eqnarray}\label{Sq1}
\mathcal{S}_{q'} (z') := \sum\limits_{k=0}^{q'-2} P_{2+k,1+k}' + z' P_{1q'}' \ .
\end{eqnarray}
Again, as for (\ref{Cq}) and (\ref{Cq1}) the approximations are finite, i.e.\ given in terms of the finite number of the matrix units and belong to the algebra of the noncommutative torus, $\mathcal{A}_\theta$.

Now we can combine the approximations (\ref{Cq}), (\ref{Cq1}), (\ref{Sq}) and (\ref{Sq1}) to produce the best approximations, $\mathbf{U}$ and $\mathbf{V}$, on the full range $P\oplus P'$ (assuming that $P'$ was rotated by the unitary $\mathbf{W}$ to become on orthogonal complement of $P$, see the discussion after (\ref{approxtotal4})):
\begin{eqnarray}\label{finalapprox}
&&\mathbf{U} := \mathcal{C}_q \oplus \mathcal{S}_{q'} (z')  \ , \nonumber \\
&&\mathbf{V} := \mathcal{S}_q (z) \oplus \mathcal{\bar{C}}_{q'}     \ .
\end{eqnarray}
Using the matrix unit algebra (\ref{matrixunits1}) it is easy to find the algebraic relations for $\mathcal{C}_q $ and $\mathcal{S}_q (z)$:
\begin{eqnarray}\label{commutators}
\mathcal{C}_q \mathcal{S}_q (z) &=& \sum\limits_{i=0}^{q-1}\sum\limits_{j=0}^{q-2}\xi^{-i}P_{i+1,i+1}P_{j+2,j+1} + z \sum\limits_{i=0}^{q-1}\xi^{-i}P_{i+1,i+1}P_{1q} =  \sum\limits_{j=0}^{q-2}\xi^{-j-1}P_{j+2,j+1} + z P_{1q} \ ,\nonumber\\
\mathcal{S}_q (z)\mathcal{C}_q &=& \sum\limits_{i=0}^{q-1}\sum\limits_{j=0}^{q-2}\xi^{-i}P_{j+2,j+1}P_{i+1,i+1} + z \sum\limits_{i=0}^{q-1}\xi^{-i}P_{1q}P_{i+1,i+1} =  \sum\limits_{j=0}^{q-2}\xi^{-j}P_{j+2,j+1} + z \xi^{-q+1} P_{1q} = \nonumber\\
&=& \xi \mathcal{C}_q \mathcal{S}_q (z) \ .
\end{eqnarray}
Because the primed objects have exactly the same definition but with positive powers of $\xi'$ (\ref{Cq1}), we immediately get
\begin{eqnarray}\label{commutatorsprimed}
\mathcal{\bar{C}}_{q'} \mathcal{S}_{q'} (z') = \xi' \mathcal{S}_{q'} (z')\mathcal{\bar{C}}_{q'} \ .
\end{eqnarray}
Combining (\ref{finalapprox}), (\ref{commutators}) and (\ref{commutatorsprimed}) we obtain the algebraic relation for $\mathbf{U}$ and $\mathbf{V}$
\begin{eqnarray}\label{commutatorUV}
\mathbf{VU} = \mathbf{\Omega} \mathbf{UV}\ ,\ \mathrm{where}\  \mathbf{\Omega}:= \xi P \oplus \xi' P' \ .
\end{eqnarray}
Though this is in not exactly the defining relation of $\mathcal{A}_\theta$ (\ref{uvcommutator}), one can see that in the limit $q,q'\rightarrow\infty$ the relation (\ref{commutatorUV}) will approximate (\ref{uvcommutator}) with any given accuracy (assuming $\xi,\xi'\rightarrow \exp(2\pi\ii\theta)$). Using this, one can show that the truncation of any element $a$ of $\mathcal{A}_\theta$ converges to $a$ in norm \cite{Landi:2003de,Landi:2004sc} (see section \ref{se:truncation} for the explicit choice of $q$ and $q'$).

One can explicitly demonstrate the finite dimensional nature of the constructed approximation by realizing the matrix units as the matrix units in $q\times q$ (or $q'\times q'$) matrix algebra\footnote{Still, remember that the approximation (\ref{finalapprox}) is realized by the elements of the full noncommutative algebra $\mathcal{A}_\theta$. What is constructed below is the isomorphism to the matrix algebra.}
\begin{eqnarray}\label{matrixunits3}
(P_{ij})_{kl} = \delta_{ki} \delta_{jl} \ .
\end{eqnarray}
Using this representation one can easily show that one obtains the relations \eqref{finalapprox1} and \eqref{CSmatrix}, with the analogous expressions for the primed objects (remember that $\mathcal{\bar{C}}_{q'}$ is constructed with $\xi'^{-1}$ instead of $\xi$).

This is the end of the explicit demonstration that the algebra generated by $\mathbf{U}$ and $\mathbf{V}$ is isomorphic to $\mathrm{Mat}_q (\mathcal{C}^\infty (\mathbb{S}^1)) \oplus \mathrm{Mat}_q (\mathcal{C}^\infty (\mathbb{S}^1))$. We call this algebra $\mathcal{A}_{qq'}$ and it is the algebra of a fuzzy torus $\mathbb{T}^2_{qq'}$. Because the matrix elements take values in $\mathcal{C}^\infty (\mathbb{S}^1)$, now both $K$-groups are isomorphic to $\mathbb{Z}\oplus\mathbb{Z}$, so the $K$-theoretical obstruction of the naive truncation has been removed.

\section*{Acknowledgements}
FL acknowledges the support of the COST action QSPACE, the INFN Iniziativa Specifica GeoSymQFT and Spanish MINECO under project MDM-2014-0369 of ICCUB (Unidad de Excelencia `Maria de
Maeztu').

\bibliographystyle{utphys}
\bibliography{actiondiracbib}

\end{document}